\documentclass[10pt, twocolumn, notitlepage, superscriptaddress, prb, longbibliography]{revtex4-2}

\usepackage{here}
\usepackage{amsmath}         
\usepackage{graphicx}
\usepackage{braket}         
\usepackage{lettrine}
\usepackage{xcolor}
\usepackage{times}
\usepackage{hhline}
\usepackage{tabularx}
\usepackage{calc}
\usepackage{soul}
\usepackage{appendix}
\usepackage{dcolumn}
\usepackage{bm}
\usepackage{float} 
\usepackage{subfigure}

\usepackage[colorlinks=true,linkcolor=blue, citecolor=blue, urlcolor=blue]{hyperref}%

\begin{document}

\preprint{APS/123-QED}

\title{Phenomenology of orbital torque, pumping and mixing conductance in metallic bilayers}

\author{Xiaobai Ning}
\affiliation{Fert Beijing Institute, School of Integrated Circuit Science and Engineering, National Key Laboratory of Spintronics, Beihang University, Beijing, China}
\affiliation{Aix-Marseille Universit\'e, CNRS, CINaM, Marseille, France}

\author{Henri Jaffrès} 
\affiliation{Laboratoire Albert Fert, CNRS, Thales, Université Paris-Saclay, 91767, Palaiseau, France}

\author{Weisheng Zhao}
\email{weisheng.zhao@buaa.edu.cn}
\affiliation{Fert Beijing Institute, School of Integrated Circuit Science and Engineering, National Key Laboratory of Spintronics, Beihang University, Beijing, China}

\author{Aur\'{e}lien Manchon}
\email{aurelien.manchon@univ-amu.fr}%
\affiliation{Aix-Marseille Universit\'e, CNRS, CINaM, Marseille, France}

\date{\today}

\begin{abstract}
The conversion between spin and orbital currents is at the origin of the orbital torque and its Onsager reciprocal, the orbital pumping. Here, we propose a phenomenological model to describe the orbital torque in magnetic bilayers composed of an orbital source (i.e., a light metal such as Ti, Ru, CuOx...) and a spin-orbit coupled magnet (i.e., typically Ni, (Co/Pt)$_n$, etc.). This approach accounts for spin-to-orbit and orbit-to-spin conversion in the ferromagnet and at the interface. We show that the orbital torque arises from a compromise between orbital current injection from the orbital source to the ferromagnet and spin current backflow from the ferromagnet back to the orbital source. We also discuss the concept of orbital-mixing conductance and introduce the "orbit-spin-" and "spin-orbit-mixing" conductances that govern the orbital torque and orbital pumping, respectively.
\end{abstract}

\maketitle


\section{INTRODUCTION}
Spin-orbit torque (SOT) induced by an electrical current has garnered significant attention for effectively manipulating the magnetization in spintronic heterostructures \cite{Gambardella2011,Manchon2019}, standing out as a promising candidate for next-generation memory and logic devices \cite{Dieny2020,Guo2021}. In a bilayer composed of a nonmagnetic metal (N) and a ferromagnet (F), it encompasses a two-step process: the generation of a spin current via spin Hall effect (SHE) \cite{Hoffmann2013,Sinova2015} or spin Rashba-Edelstein effect (SREE) \cite{Manchon2015,Bihlmayer2022} in layer N, followed by its absorption by the adjacent layer F. Because both SHE and SREE arise from spin-orbit coupling (SOC), efficient SOT devices require the use of heavy metals (Pt, W, etc.). An alternative to SOT can be achieved by using currents of orbital angular momentum rather than spin currents. Such orbital currents can be efficiently generated in light metals (Ti, Al, Zr, oxidized Cu) upon charge-orbit interconversion mechanisms such as the orbital Hall effect (OHE) \cite{Zhang2005,Go2018} and the orbital Rashba-Edelstein effect (OREE) \cite{Go2017,Go2021}. In this case, an orbital current is generated in N in the absence of SOC, and is subsequently absorbed in F through orbit-to-spin conversion mediated by SOC \cite{Go2020}. This orbital torque and its parent effect, the orbital (Hall or Rashba) magnetoresistance, have been reported recently \cite{Ding2020,Lee2021,Lee2021a,Krishnia2023,Yang2024}. A hallmark of the orbital torque is its dependence as a function of F's thickness. Since the orbital current is more likely to perform long-range transport than the spin current in F \cite{Go2023}, the strength of the torque exhibits a steady enhancement with increasing F's thickness, while the conventional SOT quickly reaches saturation \cite{Ding2022b,Ding2024,Hayashi2023,Bose2023,Krishnia2023,Yang2024}. 

A practical way to model SOT in an N/F bilayer is to combine a drift-diffusion model that describes the generation of the spin current in N, and the spin-mixing conductance boundary condition that describes the spin current absorption in F \cite{Tserkovnyak2002, Haney2013}. This approach is valid only when the SOC of both F and the N/F interface is neglected. For orbital torque, the charge-to-orbital conversion has been computed from first principles \cite{Tanaka2008, Kontani2009, Salemi2022, Pezo2022, Go2024}, and the resulting torque in N/F bilayers have been obtained using, e.g., tight-binding modeling \cite{Go2020}. Such an approach provides an instructive microscopic description but lacks transparency. Therefore, a phenomenological approach similar to the one used to model SOT is highly desired. An attempt has been made to extend the spin drift-diffusion theory to orbital transport \cite{Sala2022,Krishnia2023} by introducing a coupling length representing the conversion between spin and orbital moments in N, while the absorption of the spin and orbital currents is modeled by an effective mixing conductance at the N/F interface. This approach provides reasonable scaling but assumes that the orbital current is fully absorbed at the interface, which is incorrect as pointed out by Ref. \cite{Go2023}: since the orbital moment does not directly couple to the magnetization, it does not experience dephasing, in contrast with the spin moment. Recent microscopic theories \cite{Han2022,Ning2024} have pointed out additional features that enrich the physics of orbital transport. First, since the orbital momentum operator does not commute with the Hamiltonian in general, it is not a conserved quantity (even in the absence of SOC). This has two important consequences: (i) the orbital moment exhibits a much smaller diffusion coefficient than the spin or the charge \cite{Ning2024}. In addition, an orbital moment penetrating a normal metal experiences a precession around the local crystal field \cite{Han2022}, similar to the precession of a spin current penetrating a ferromagnet. Second, in the presence of SOC, spin and orbital currents are coupled through a so-called spin-to-orbit and orbit-to-spin polarization \cite{Ning2024}. This coefficient is different from the spin-orbit length introduced in Ref. \cite{Sala2022}, as further discussed in the present Article. Although additional microscopic investigations are necessary to clarify the nature of these effects at interfaces and the impact of disorder and defects, these features are remarkable and must be accounted for to properly model orbital transport. 

In the present work, we develop a drift-diffusion model that describes the spin and orbital transport induced by SHE and OHE in N/F bilayers. This approach accounts for spin and orbital diffusion inside F, as well as spin-to-orbit conversion in F and at the N/F interface. It enables us to clarify the parameters that control the orbital torque, demonstrating the interplay between orbital current injection and spin current backflow. Finally, we discuss the concept of orbital-mixing conductance and its connection to orbital pumping.

\section{THEORY}

We intend to describe the phenomenology of orbital torque in an N/F bilayer, as depicted in Fig. \ref{mechanism}. The theory of spin transfer torque has been studied in great detail over the past two decades, pointing out the central role of spin precession and dephasing in F \cite{Stiles2002,Zhang2002,Petitjean2012,Lim2021}. Since the spin dephasing length is usually short in most transition metal ferromagnets (typically, $\approx 1$ nm), the physics governing the absorption of an incoming spin current in a ferromagnet is often modeled by the so-called interfacial spin-mixing conductance \cite{Brataas2001,Zwierzycki2005,Brataas2006}, i.e., a boundary condition that connects the interfacial spin current with the drop in spin chemical potential across the interface. It is tempting to develop a similar concept for orbital torque, the "orbital-mixing conductance", that connects the impinging orbital current with the drop in orbital chemical potential. Nonetheless, since orbital currents diffuse over much longer distances than spin currents in ferromagnets \cite{Ding2022b,Ding2024,Hayashi2023,Bose2023,Krishnia2023,Yang2024}, a comprehensive model needs to properly account for the actual spin precession and dephasing inside F. 

\subsection{Two-channel Spin-Orbit Transport\label{s:model}}

We assume that N accommodates both OHE and SHE. The spin diffusion equations in the presence of both SOC and magnetic exchange have been studied intensively over the past two decades, resulting in rather complex intermingling of intrinsic and extrinsic SHE, spin swapping, spin precession, dephasing, and relaxation \cite{Shchelushkin2005,Petitjean2012,Pauyac2018}. To keep our theory tractable, we chose to retain only the terms that are the most relevant to the phenomena we are interested in. Assuming that in a nonmagnetic metal, the spin and orbital transport are governed by diffusion, relaxation, SHE, and OHE, the diffusion equations that connect the spin (orbital) chemical potential ${\bm\mu}_{s(o)}^{\rm N}$ to the spin (orbital) current tensor ${\bf J}_{s(o)}^{\rm N}$ read \cite{Shchelushkin2005}
\begin{eqnarray}
{\bf J}_s^{\rm N}&=&-\sigma_s^{\rm N}\partial_z{\bm\mu}_s^{\rm N}+\sigma_{\rm SHE}\hat{\bm\sigma}\times {\bf E},\;\partial_z^2{\bm\mu}_s^{\rm N}=\frac{{\bm\mu}_s^{\rm N}}{{\lambda_s^{\rm N}}^2} \label{spin-diff},\\
{\bf J}_o^{\rm N}&=&-\sigma_o^{\rm N}\partial_z{\bm\mu}_o^{\rm N}+\sigma_{\rm OHE}\hat{\bm l}\times {\bf E},\;\partial_z^2{\bm\mu}_o^{\rm N}=\frac{{\bm\mu}_o^{\rm N}}{{\lambda_o^{\rm N}}^2}. \label{orbital-diff}
\end{eqnarray}
Here $\hat{\bm\sigma}$ ($\hat{\bm l}$) is the unit vector of the spin (orbital) polarization, $\lambda_{s(o)}^{\rm N}$ is the spin (orbital) relaxation length, $\sigma_{s(o)}^{\rm N}$ is the spin (orbital) longitudinal conductivity and $\sigma_{\rm SHE(OHE)}$ is the spin (orbital) Hall conductivity. In other words, we consider that the spin and orbital degrees of freedom propagate along distinct channels. Notice that, without loss of generality, the spin and orbital currents are expressed in the unit of a charge current to avoid unnecessarily cumbersome notation.

When SOC is present, the two channels are coupled, and thus, an incoming spin current is accompanied by an orbital current and vice-versa \cite{Ning2024}. We disregard this effect in Eqs. \eqref{spin-diff}-\eqref{orbital-diff} because it does not add new physics and only leads to renormalized transport coefficients. Nonetheless, in F, this effect is crucial to convert the incoming orbital current into a spin current and drive the orbital torque. Therefore, the drift-diffusion equations can be modified as
\begin{align}
    \boldsymbol{J}_s^{\rm F}&=-\sigma_s^{\rm F}\partial_z\boldsymbol{\mu}_s^{\rm F}-P_{os}\sigma_o^{\rm F}\partial_z\boldsymbol{\mu}_o^{\rm F},\label{precession_bulk_equation7}\\
    \boldsymbol{J}_o^{\rm F}&=-\sigma_o^{\rm F}\partial_z\boldsymbol{\mu}_o^{\rm F}-P_{so}\sigma_s^{\rm F}\partial_z\boldsymbol{\mu}_s^{\rm F},\label{precession_bulk_equation8}
\end{align}
where $P_{os}$ and $P_{so}$ stand for the orbit-to-spin and spin-to-orbit polarization, respectively. In Ref. \cite{Ning2024}, we have computed these coefficients in realistic materials and demonstrated that the spin-to-orbit and orbit-to-spin polarizations can be as large as spin polarization in transition metals (i.e., several tens of percent). In addition, the itinerant spin density directly couples to the magnetization via the exchange coupling, so that the spin chemical potential obeys \cite{Petitjean2012,Pauyac2018,Lim2021}
\begin{equation}
\left(\partial_z^2-\frac{1}{\lambda_{sf}^2}\right)\boldsymbol{\mu}_s^{\rm F}-\frac{1}{\lambda_J^2}\boldsymbol{\mu}_s^{\rm F}\times{\bf m}-\frac{1}{\lambda_{\phi}^2}{\bf m}\times(\boldsymbol{\mu}_s^{\rm F}\times{\bf m})=0, \label{precession_bulk_equation5}
\end{equation}
where $\lambda_{sf}$ is the spin diffusion length, $\lambda_J$ is the spin precession length and $\lambda_{\phi}$ is the spin dephasing length. Hence, the component of the spin density that is aligned on the magnetization ${\bf m}$ simply diffuses into F, whereas the transverse components experience a damped precession. In other words, these transverse components have the form ${\bm \mu}_s^{F}={\bf A}e^{(\alpha-i\beta)z}+{\bf B}e^{-(\alpha-i\beta)z}$, with 
\begin{align}
    \begin{aligned}
        \alpha=\frac{1}{2}\left[\sqrt{\left(\frac{1}{\lambda_{sf}^2}+\frac{1}{\lambda_{\phi}^2}\right)^2+\frac{1}{\lambda_{J}^4}}+\frac{1}{\lambda_{sf}^2}+\frac{1}{\lambda_{\phi}^2}\right]^{\frac{1}{2}},\\
        \beta=\frac{1}{2}\left[\sqrt{\left(\frac{1}{\lambda_{sf}^2}+\frac{1}{\lambda_{\phi}^2}\right)^2+\frac{1}{\lambda_{J}^4}}-\frac{1}{\lambda_{sf}^2}-\frac{1}{\lambda_{\phi}^2}\right]^{\frac{1}{2}}. \label{precession_bulk_general6}
    \end{aligned}
\end{align}

In contrast, the orbital density only couples to the crystal field potential, which induces its relaxation \cite{Sohn2024}. The drift-diffusion equation is therefore
\begin{align}
\partial_z^2{\bm\mu}_o^{\rm F}&=\frac{{\bm\mu}_o^{\rm F}}{{\lambda_o^{\rm F}}^2}.
\end{align}
This expression is the simplest phenomenological form of the orbital diffusion and neglects the precession of the orbital momentum around the crystal field predicted in Refs. \cite{Han2022,Ning2024} and that remained to be thoroughly discussed. 

\subsection{Boundary Conditions}

At the N/F interface, (spin and orbital) currents and chemical potentials are connected by a set of conductances, akin to Ohm's law, 
\begin{align}
    \boldsymbol{J}_{s}^{\rm F}=\boldsymbol{J}_{s}^{\rm N}=G_s(\boldsymbol{\mu}_s^{\rm N}-\boldsymbol{\mu}_{s}^{\rm F})+G_{os}(\boldsymbol{\mu}_o^{\rm N}-\boldsymbol{\mu}_o^{\rm F}),\label{precession_interfacial_boundary5}\\
    \boldsymbol{J}_o^{\rm F}=\boldsymbol{J}_o^{\rm N}=G_o(\boldsymbol{\mu}_o^{\rm N}-\boldsymbol{\mu}_o^{\rm F})+G_{so}(\boldsymbol{\mu}_s^{\rm N}-\boldsymbol{\mu}_s^{\rm F}),\label{precession_interfacial_boundary6}
\end{align}
where $G_{s(o)}$ is the interfacial spin (orbital) conductance, and $G_{os(so)}$ model the transfer between spin and orbital momenta that may occur in the presence of interfacial SOC. These interfacial spin-orbit conductances play the same role as the spin-orbit polarizations introduced above: they unlock the conversion of an incoming orbital current into a spin current and enable orbital torque. The coefficient $G_{os}$ is of particular interest to us as it connects the orbital chemical potential to the spin current and acts like an orbital-spin mixing conductance, as discussed further below. Notice that, in principle, these interfacial conductances are $3\times3$ tensors as spin and orbital precession might take place at the interface. Understanding these effects would require systematic {\em ab initio} treatment that goes beyond the scope of the present work.

As mentioned above, the spin precession and dephasing taking place in F usually occur over a short distance, typically $\sim$1 nm. Therefore, it is often convenient to assume that the spin density injected inside it is immediately absorbed close to the interface, i.e., $\boldsymbol{J}_{s}^{\rm F},\boldsymbol{\mu}_{s}^{\rm F}\rightarrow0$. In this case, in the limit of FM thickness larger than the typical spin decoherence length, the boundary conditions are reduced to 
\begin{align}
    {\bf J}_s^{\rm N}&=G_{s,r}^{\uparrow\downarrow}{\bf m}\times({\bm\mu}_s^{\rm N}\times{\bf m})+G_{s,i}^{\uparrow\downarrow}{\bf m}\times{\bm\mu}_s^{\rm N}+G_{os}({\bm\mu}_o^{\rm N}-{\bm\mu}_o^{\rm F}),\label{interface_orbital_boundary4}\\
     {\bf J}_o^{\rm F}&={\bf J}_o^{\rm N}=G_o({\bm\mu}_o^{\rm N}-{\bm\mu}_o^{\rm F})+G_{so}{\bm\mu}_s^{\rm N}, \label{interface_orbital_boundary5}
\end{align}
where $G^{\uparrow\downarrow}_s=G^{\uparrow\downarrow}_{s,r}+iG^{\uparrow\downarrow}_{s,i}$ is the spin-mixing conductance \cite{Brataas2001} (the factor "2" appearing in the original definition is absorbed in $G^{\uparrow\downarrow}_s$). In the next section, we solve this set of equations to compute the orbital torque in N/F bilayers.

\section{ANATOMY OF ORBITAL TORQUE}
\subsection{Definition of the SOT}
In the following, we assume that the torque originates from the transfer of spin angular momentum from the itinerant electrons to the magnetization. In other words, we assume that the orbital current does not directly participate in the SOT \cite{Go2020a}. As a result, the torque ${\bm\tau}$ reads 
\begin{equation}
        {\bm\tau}=-\int_{0}^{d_{\rm F}}{\bf m}\times({\bm\nabla}\cdot{\bf J}_s\times{\bf m})={\bf m}\times({\bf J}_s^{\rm N}(0)\times{\bf m}), \label{interface_spin_torque1}
\end{equation}
where ${\bf J}_s^{\rm N}(0)$ is the spin current at the N/F interface. In other words, the SOT strength is solely characterized by the interfacial spin current, regardless of its origin (SHE or OHE). In the configuration we adopt, the bilayer is extended along the $z$-direction, and the electric field is applied along ${\bf x}$. As a result, SHE and OHE generate a current of angular momentum polarized along ${\bf y}$. According to Eq. \eqref{precession_bulk_equation5}, the penetration of the spin into F is accompanied by a precession, and consequently, the SOT possesses two components, ${\bm\tau}=\tau_{\parallel}{\bf m}\times({\bf y}\times {\bf m})+\tau_\bot{\bf m}\times {\bf y}$, usually referred to as the in-plane and perpendicular components. The former is responsible for the damping-like torque and the latter for the field-like torque \cite{Manchon2019}. In the following, we express the torque as a complex quantity, defined $\tau=\tau_{\parallel}+i\tau_\bot$. With this formulation, the SOT takes the general form
 \begin{align}
    \tau_{\rm SHE}=\eta_s\sigma_{\rm SHE}E\left(1-\cosh^{-1}\frac{d_{\rm N}}{\lambda_s^{\rm N}}\right),\label{spin_efficiency}\\
    \tau_{\rm OHE}=\eta_o\sigma_{\rm OHE}E\left(1-\cosh^{-1}\frac{d_{\rm N}}{\lambda_o^{\rm N}}\right),\label{orbital_efficiency}
\end{align}
where $\eta_s$ and $\eta_o$ are the (complex) spin and orbital transmission efficiencies, and $d_{\rm N}$ is the thickness of layer N. The detailed expression of these efficiencies is rather complex and difficult to interpret transparently since they involve the interplay between spin and orbital diffusion, interfacial scattering, and spin-orbit interconversion. Therefore, before discussing the numerical results, it is worth inspecting the general form of these two transmission efficiencies in certain limits. To do so, we define the effective (spin and orbital) conductance $\widetilde{G}_{s(o)}^{\rm N(F)}$ that quantifies the ability for a given metal (N or F) to absorb the (spin or orbital) current,
\begin{equation}
    \widetilde{G}_{s(o)}^{\rm N(F)}=\frac{\sigma_{s(o)}^{\rm N(F)}}{\widetilde{\lambda}_{s(o)}^{\rm N(F)}},\label{conductance}
\end{equation}
where the (possibly complex) spin (orbital) diffusion lengths read $\widetilde{\lambda}_s^{\rm F}=(\alpha-i\beta)^{-1}$ and $\widetilde{\lambda}_{s(o)}^{\rm N(F)}=\lambda_{s(o)}^{\rm N(F)}\coth\left[d_{N(F)}/\lambda_{s(o)}^{\rm N(F)}\right]$. The larger the conductivity and the smaller the diffusion length, the larger the ability of the metal to absorb a (spin or orbital) current. In addition, we assume that the thickness of F is larger than the spin dephasing length, $e^{2d_{\rm F}(\alpha-i\beta)}\gg1$, and we account for SOC only perturbatively so that the interfacial spin-to-orbit (orbit-to-spin) polarization is $\gamma_{so(os)}=G_{os(so)}/G_{o(s)}\ll1$. These limits, physically reasonable, allow us to determine explicit expressions of the spin and orbital transmission efficiencies.

\begin{figure}[!t] 
\centering 
\includegraphics[width=0.48\textwidth]{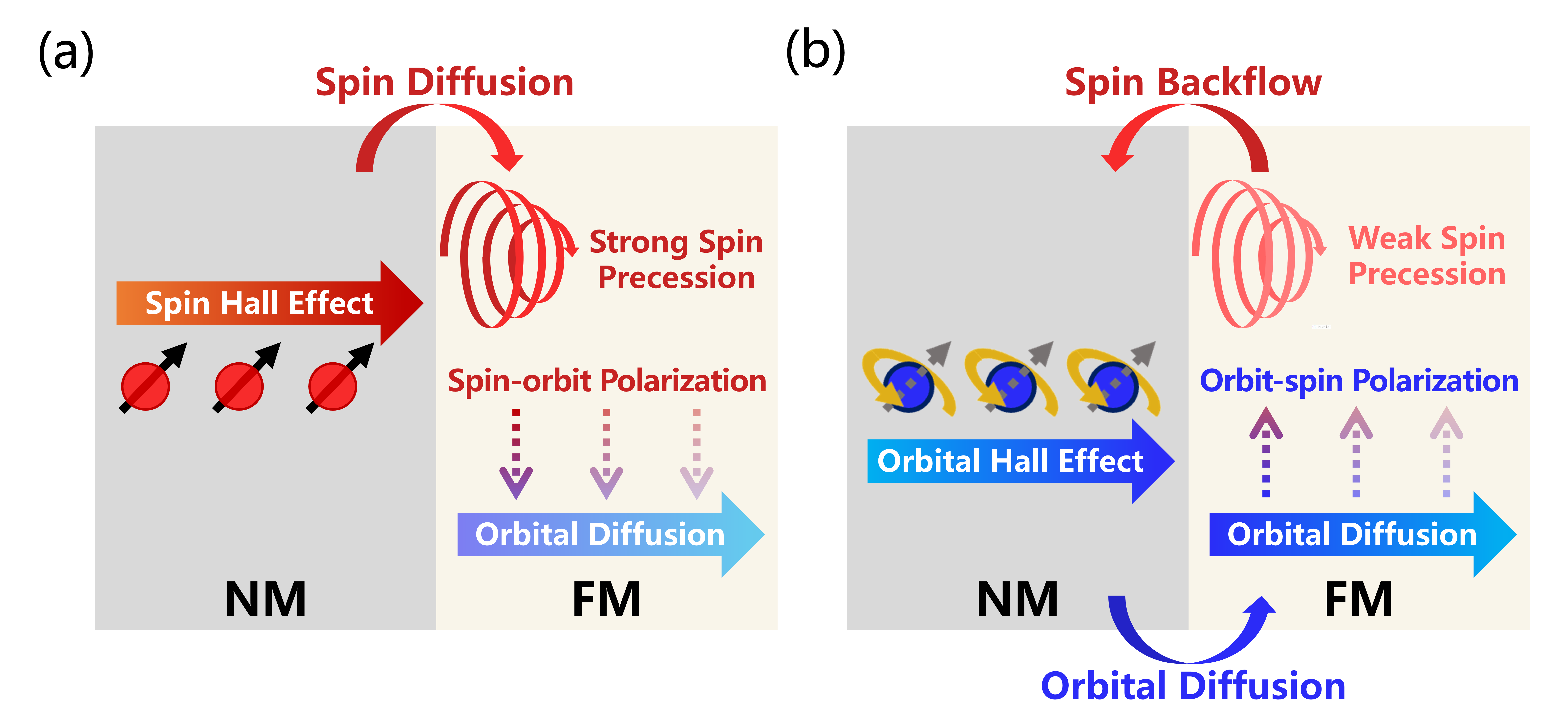}
\caption{(Color online) Current intensities and diffusion directions for both spins and orbits when (a) SHE or (b) OHE is dominant in N.} 
\label{mechanism} 
\end{figure}

\subsection{Spin Hall effect-driven SOT}

To clarify the physics behind the orbital torque, let us first consider the standard case of the SHE-driven torque ($\sigma_{\rm SHE}\neq0,\;\sigma_{\rm OHE}=0$). Considering first the diffusion of the spin current into F, Eq. \eqref{precession_bulk_equation5}, the spin transmission efficiency reads
\begin{equation}
    \eta_s=\left(1+\frac{\widetilde{G}_s^{\rm N}}{\widetilde{G}_s^{\rm F}}\right)^{-1}.\label{efficiency_s_bulk}
\end{equation}
The SOT efficiency is determined by the relative strengths between the effective spin conductances of N and F. For example, if N is highly conductive, the spin current is likely to stay in N instead of penetrating F, reducing the proportion of transmitted spin current and, thus, the SOT. In other words, to maximize the SHE-driven SOT, the bilayer should be designed such that $\widetilde{G}_s^{\rm F}\gg\widetilde{G}_s^{\rm N}$.  If SOC is turned on in F, the spin current is accompanied by an orbital current, see Eqs. \eqref{precession_bulk_equation7}-\eqref{precession_bulk_equation8}. The contribution of this orbital current to the SOT is indirect as it simply renormalizes the efficiency by a factor $1+\gamma_{so}\gamma_{os}\approx 1$. This negligible contribution of the orbital currents is rather obvious since the primary mechanism for SOT is the SHE. When the complete absorption of the spin current at the interface is assumed, Eq. \eqref{efficiency_s_bulk} remains valid with $\widetilde{G}_s^{\rm F}\rightarrow G_s^{\uparrow\downarrow}$. The spin mixing conductance $G_{\uparrow\downarrow}$ includes the non-collinear spin precession contributions, so the resultant $\eta_s$ can have both real and imaginary parts, corresponding to damping-like torque and field-like torque, respectively.

\begin{figure*}[!t] 
\centering 
\includegraphics[width=\textwidth]{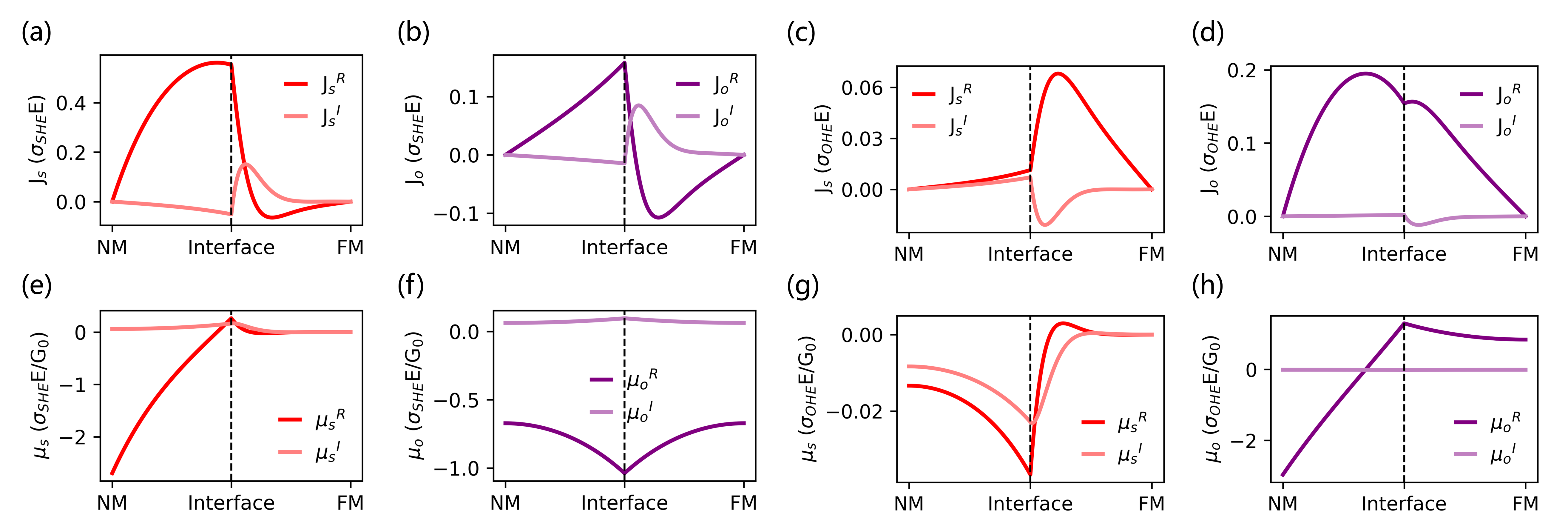}
\caption{(Color online) (a, c) Spin and (b, d) orbital current profiles induced by SHE (a, b, e, f) and OHE (c, d, g, h). The corresponding spin and orbital chemical potential profiles are shown in (e, g) and (f, h), respectively. In the definition of the chemical potentials, we introduce the scaling parameter $G_0=10^{15} \Omega^{-1}\cdot m^{-2}$. The superscript $^R$ and $^I$ refer to the real [$\propto {\bf m}\times({\bf y}\times{\bf m})$] and imaginary parts [$\propto {\bf m}\times {\bf y}$] of the spin and orbital vectors. The material parameters are set to $d_{\rm N}=d_{\rm F}=5nm$, $\lambda_s^{\rm N}=\lambda_s^{\rm F}=3nm$, $\lambda_o^{\rm N}=\lambda_o^{\rm F}=5nm$, $\lambda_{\phi}=1nm$, $\lambda_J=0.5nm$, and $\sigma_s^{\rm N}=\sigma_s^{\rm F}=\sigma_o^{\rm N}=\sigma_o^{\rm F}=10^6$ $\Omega^{-1}\cdot$m$^{-1}$.}
\label{profileshe}
\end{figure*}

\subsection{Orbital Hall effect-driven SOT}

The orbital torque efficiency, $\eta_o$, presents a very different structure that manifests the alternative scenario governing orbital injection. Assuming that spin-to-orbit conversion takes place in F [Eqs. \eqref{precession_bulk_equation7}-\eqref{precession_bulk_equation8}], we obtain the efficiency
\begin{equation}
    \eta_o=P_{os}\left(1+\frac{\widetilde{G}_s^{\rm F}}{\widetilde{G}_s^{\rm N}}\right)^{-1}\left(1+\frac{\widetilde{G}_o^{\rm N}}{\widetilde{G}_o^{\rm F}}\right)^{-1}. \label{efficiency_o_bulk}
\end{equation}
The efficiency is the product of three terms. First, the term $\sim\left(1+\frac{\widetilde{G}_o^{\rm N}}{\widetilde{G}_o^{\rm F}}\right)^{-1}$ represents the injection of the orbital current from N into F. Once in F, this orbital current is converted into a spin current via the orbit-to-spin polarization factor $P_{os}$. Finally, the last term $\sim\left(1+\frac{\widetilde{G}_s^{\rm F}}{\widetilde{G}_s^{\rm N}}\right)^{-1}$ represents the injection of the spin current from F into N. In other words, since the SOT is due to the absorption of the spin current, it is directly related to the spin current {\em backflow} from F to N, as illustrated in Fig. \ref{mechanism}(b). This is a distinctive feature compared to SHE-driven SOT that scales with the direct injection of the spin current from N to F [see Eq. \eqref{efficiency_s_bulk}].

Equation \eqref{efficiency_o_bulk} corresponds to a scenario where the orbit-to-spin conversion takes place in F, as achieved in Refs. \cite{Sala2022,Hayashi2023,Bose2023}. Alternatively, the orbit-to-spin conversion can be boosted by inserting a thin layer with strong SOC between N and F, e.g., CuOx/Pt/TmG \cite{Ding2020}, Cr/(Gd,Pt)/Co \cite{Lee2021a} or Co/Pt/CuOx \cite{Krishnia2024}. In this case, the conversion rather happens at the N/F interface, Eqs. \eqref{precession_interfacial_boundary5}-\eqref{precession_interfacial_boundary6}, and the expression of $\eta_o$ changes to
\begin{eqnarray}
\eta_o=\gamma_{os}\frac{\widetilde{G}_s^{\rm F}}{G_s}\left(1+\frac{\widetilde{G}_s^{\rm F}}{\widetilde{G}_s^{\rm N}}\right)^{-1}\left[1+\widetilde{G}_o^{\rm N}\left(\frac{1}{G_o}+\frac{1}{\widetilde{G}_o^{\rm F}}\right)\right]^{-1}.\label{efficiency_o_interface}
\end{eqnarray}
As depicted in Fig. \ref{mechanism}(b), it contains the orbital transmission, $G_o$, the orbit-to-spin conversion, $G_{os}$, the spin precession, $\widetilde{G}_s^{\rm F}$, and the spin backflow, $G_s$. If one assumes that the spin current is absorbed at the interface, we simply need to make the substitution $\widetilde{G}_s^{\rm F},G_s\rightarrow G_s^{\uparrow\downarrow}$. 

These different expressions are instrumental in determining the scenario of OHE-driven SOT. Importantly, using different assumptions, such as bulk versus interfacial orbit-to-spin conversion and interfacial absorption of the incoming spin current (i.e., adopting the spin-mixing conductance approach), does not modify the overall picture. To optimize the OHE-driven SOT, one needs to maximize the injection of the orbital current from N into F and maximize the spin current backflow from F into N. So ideally, the materials composing the bilayer should be such that $\widetilde{G}_s^{\rm N}\gg\widetilde{G}_s^{\rm F}$ and $\widetilde{G}_o^{\rm F}\gg\widetilde{G}_o^{\rm N}$.

\section{ORBITAL TORQUE MODELING}

\subsection{Current and Chemical Potential Profiles}

The expressions of the SOT discussed in the previous section were obtained assuming thick F and weak SOC. Let us now inspect the SOT efficiency by computing the orbital and spin currents across the N/F bilayer using the drift-diffusion model exposed in Section \ref{s:model}. For these calculations, we account for the spin precession inside F, Eq. \eqref{precession_bulk_equation5}.  As a benchmark, we first consider the case of SHE-driven SOT and we reproduce the (spin and orbital) current and density profiles in Fig. \ref{profileshe}. The spin current and density profile shown in Fig. \ref{profileshe}(a) and (e) are well-known and have been discussed elsewhere (see, e.g., Ref. \cite{Haney2013}). Since spin-to-orbit conversion takes place in F, the penetration of the spin current is accompanied by an orbital current [Fig. \ref{profileshe}(b)] that simply follows the spin profile. Notice that the orbital density diffuses over much longer distances. This orbital current only weakly affects the SOT, as discussed in more detail below.

Let us now pay attention to the spin and orbital current profiles induced by OHE, displayed in Fig. \ref{profileshe}(c,d). Since the transport is driven by OHE in N, the spin current flowing in N is negligible [Fig. \ref{profileshe}(c,g)], whereas the orbital current displays the conventional exponential profile reflecting the Hall effect [Fig. \ref{profileshe}(d,h)] (compare with [Fig. \ref{profileshe}(a,c)]). Once in F, where orbit-to-spin conversion occurs, the spin current is quickly absorbed close to the interface, experiencing a damped oscillation [Fig. \ref{profileshe}(c,g)], whereas the orbital current slowly diffuses [Fig. \ref{profileshe}(d,h)].

Finally, we stress that the nonequilibrium spin and orbital moments (current and density) possess two components, in-plane [$\propto {\bf m}\times({\bf y}\times{\bf m})$] and perpendicular to the (${\bf m},{\bf y}$) plane  [$\propto{\bf m}\times {\bf y}$], and represented respectively by the real and imaginary parts of $J_{s(o)}$ and $\mu_{s(o)}$. Fig. \ref{profileshe}(d) shows that OHE first generates an orbital current polarized along ${\bf m}\times({\bf y}\times{\bf m})$, which remains dominant during the full diffusion and orbit-to-spin conversion process. The orbital-to-spin conversion in F results in a spin current polarized along the same direction, $\propto{\bf m}\times({\bf y}\times{\bf m})$. Due to exchange coupling, this newly generated spin current precesses around the magnetization, inducing a perpendicular component, $\propto{\bf m}\times {\bf y}$ [see Fig. \ref{profileshe}(c)]. Subsequently, a perpendicular component of orbital current is generated by spin-to-orbit conversion  [see Fig. \ref{profileshe}(d)]. Since this last process involves spin-to-orbit and orbit-to-spin conversions, the overall magnitude of the perpendicular moment of the orbital current is tiny.

\subsection{Thickness Dependence of Orbital Torque}

\begin{figure}[!t] 
\centering 
\includegraphics[width=0.46\textwidth]{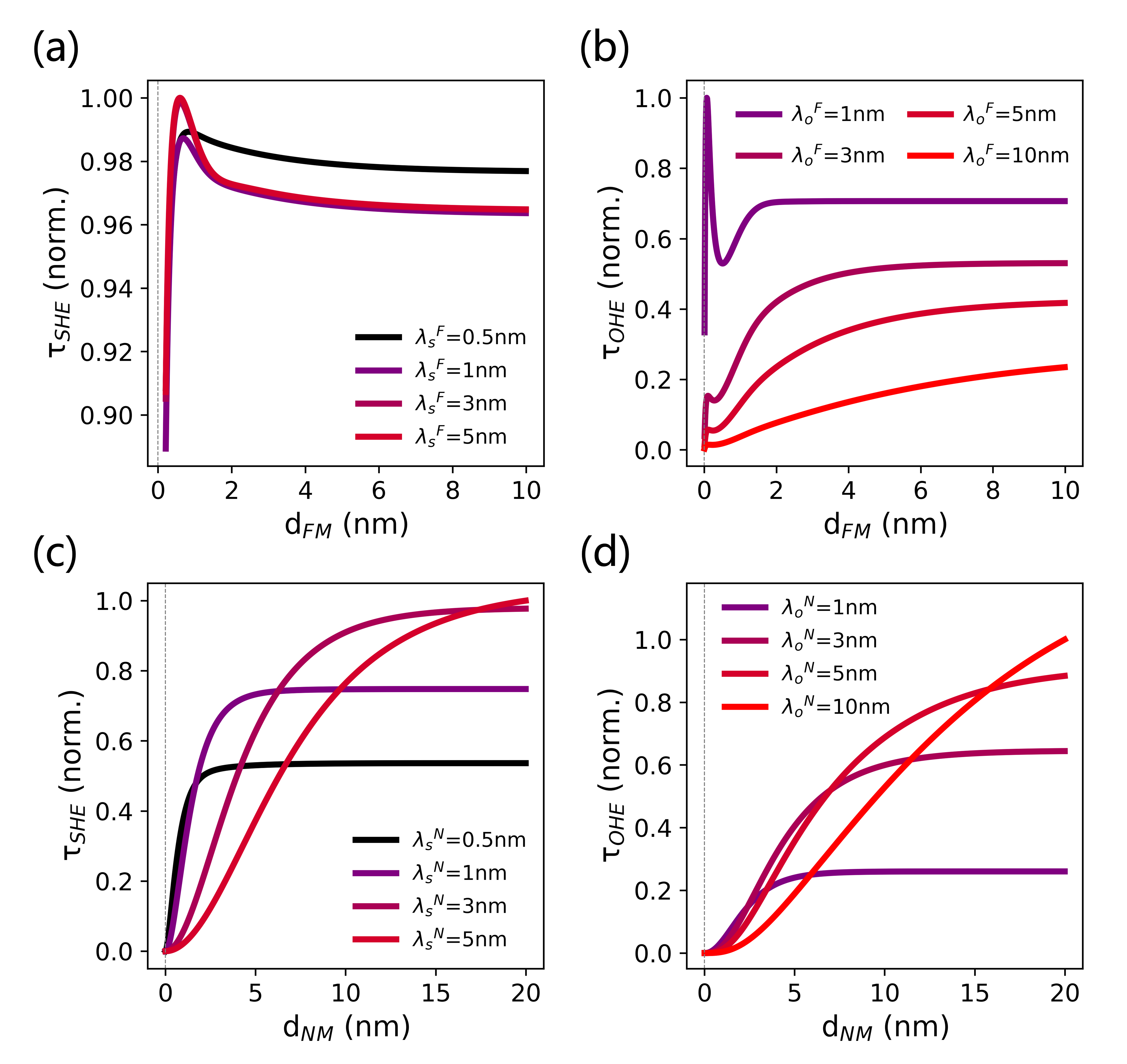}
\caption{(Color online) Damping-like SOT initiated by (a) SHE and (b) OHE as a function of F's thickness. The counterparts of N are shown in (c) and (d) accordingly. The spin-to-orbit polarization $P_{so}$ and orbit-to-spin polarization $P_{os}$ are assigned a value of 0.5, while the remaining parameters are consistent with Fig. \ref{profileshe}.} 
\label{torque} 
\end{figure}

A hallmark of the orbital contribution to damping-like SOT is the dependence of the torque efficiency on the thickness of F \cite{Ding2022b,Ding2024,Hayashi2023,Bose2023,Krishnia2023,Yang2024}. Figures \ref{torque}(a,c) correspond to SHE-driven SOT and Fig. \ref{torque}(b,d)  correspond to OHE-driven SOT. In Fig. \ref{torque}(a), the SHE-driven SOT rapidly saturates with the thickness of F, and its magnitude is nearly independent of the spin diffusion length. This is because the spin current absorption $\sim\alpha-i\beta$ is dominantly determined by the spin precession $\lambda_J$ and spin dephasing $\lambda_{\phi}$ lengths close to the interface \cite{Haney2013}. On the contrary, the OHE-driven SOT [Fig. \ref{torque}(b)] shows a clear dependence on the thickness of F and its orbital diffusion length, as the orbital moment does directly couple to the magnetization. It is important to note that the orbital torque decreases when increasing $\lambda_o^{\rm F}$, which seems counterintuitive. In fact, the weak orbital relaxation in F enables the orbital backflow into N, thereby reducing the amount of orbital current effectively injected into F.

The dependence of SOT on the thickness of layer N is reported in Figs. \ref{torque}(c,d). Here, as we consider the total SOT, the accumulation term $(1-\cosh^{-1}d_{\rm N}/\lambda_{s(o)}^{\rm N})$ and the efficiency term $\eta_{s(o)}$ are in competition with each other, leading to different correlations between the SOT and $\lambda_{s(o)}^{\rm N}$ depending on the thickness. The accumulation term increases with $d_{\rm N}$ and saturates for $d_{\rm N}\approx 5\lambda_{s(o)}^{\rm N}$, whereas the efficiency $\eta_{s(o)}$ increases with $d_{\rm N}/\lambda_{s(o)}^{\rm N}$. In other words, the orbital torque can be optimized by choosing a short $\lambda_{o}^{\rm F}$, a long $\lambda_{o}^{\rm N}$ and $d_{\rm N}\approx 5\lambda_{o}^{\rm N}$.

\subsection{Orbital Mixing Conductance and Orbital Pumping}

We conclude this work by commenting on the nature of the orbital-mixing conductance and connecting it to orbital pumping. As mentioned in the introduction, given the tremendous success of the concept of spin-mixing conductance, it is appealing to define an orbital version of this interfacial boundary condition. The spin-mixing conductance, which connects the interfacial spin chemical potential to the interfacial spin current, has been originally defined using the scattering-matrix theory, $G_s^{\uparrow\downarrow}=(2e^2/hA)\sum_{nm}( \delta_{nm}-r^{\uparrow}_{mn}r^{\downarrow*}_{mn})$, where $r_{nm}^\sigma$ is the spin-dependent reflection coefficient for the $n,m$ modes at the N/F interface \cite{Brataas2001,Zwierzycki2005,Brataas2006}. In the framework of the drift-diffusion theory developed here, we obtain the alternative definition
\begin{eqnarray}
    G_s^{\uparrow\downarrow}=\left[\frac{1}{G_s}+\frac{1}{\sigma_s^{\rm F}(\alpha-i\beta)}\right]^{-1},\label{conductance_spin}
\end{eqnarray}
which expresses the spin-mixing conductance as a function of the spin precession, dephasing, and relaxation lengths in F. 

One can conceive different definitions for the orbital-mixing conductance. The definition that is the closest to the spin-mixing conductance connects the interfacial {\em orbital} chemical potential to the interfacial {\em orbital} current, $G_o$ [see Eq. \eqref{precession_interfacial_boundary6}]. However, because the orbital current is absorbed far away from the interface and does not directly produce a torque on the magnetization, this definition is not particularly useful for modeling SOT. A more practical definition connects the interfacial {\em orbital} chemical potential to the interfacial {\em spin} current that is quickly absorbed close to the interface, as discussed in the previous section. The definition of this "orbit-spin-mixing conductance" differs depending on the adopted model. If the orbit-to-spin conversion is accomplished inside F, we obtain
\begin{equation}
    G_{os}^m=G_oP_{os}. \label{conductance_bulk}
\end{equation}
The orbit-spin-mixing conductance arises from the transmission of orbital currents at the N/F interface followed by the orbit-to-spin conversion in F. It is worth noticing that no spin precession terms appear in Eq. \eqref{conductance_bulk} since the spin current is simply dragged by the orbital current in F, as reported in Fig. \ref{profileshe}(b). If the orbit-to-spin conversion occurs at the interface, we obtain
\begin{equation}
    G_{os}^m=\frac{G_{os}}{G_s}\sigma_s^{\rm F}(\alpha-i\beta). \label{conductance_interface}
\end{equation}
Likewise, we can attribute the orbit-spin-mixing conductance to three sub-steps: the orbit-to-spin conversion by $G_{os}$, the spin precession by $\sigma_s^{\rm F}(\alpha-i\beta)$, and the spin backflow by $G_s$. It is distinct from the bulk orbit-to-spin conversion scenario, as the spin current is not dragged by the orbital current but arises from interfacial "filtering"; as a consequence, the spin current backflow controls the magnitude of the orbit-spin-mixing conductance. 

Figure \ref{conductance} displays the dependence of the (complex) spin-mixing conductance (red lines) and the orbit-spin-mixing conductance (purple) as a function of (a) the interfacial spin conductance and (b) the spin precession length. Compared to the spin-mixing conductance, the orbit-spin-mixing conductance is generally more sensitive to the changes in both interfacial and ferromagnetic parameters. It is also worth noticing that the spin-mixing conductance and the orbit-spin-mixing conductance show different signs of correlation with the interfacial spin conductance, reflecting the diffusion and backflow of spin as illustrated in Fig. \ref{mechanism}.

\begin{figure}[!t] 
\centering 
\includegraphics[width=0.46\textwidth]{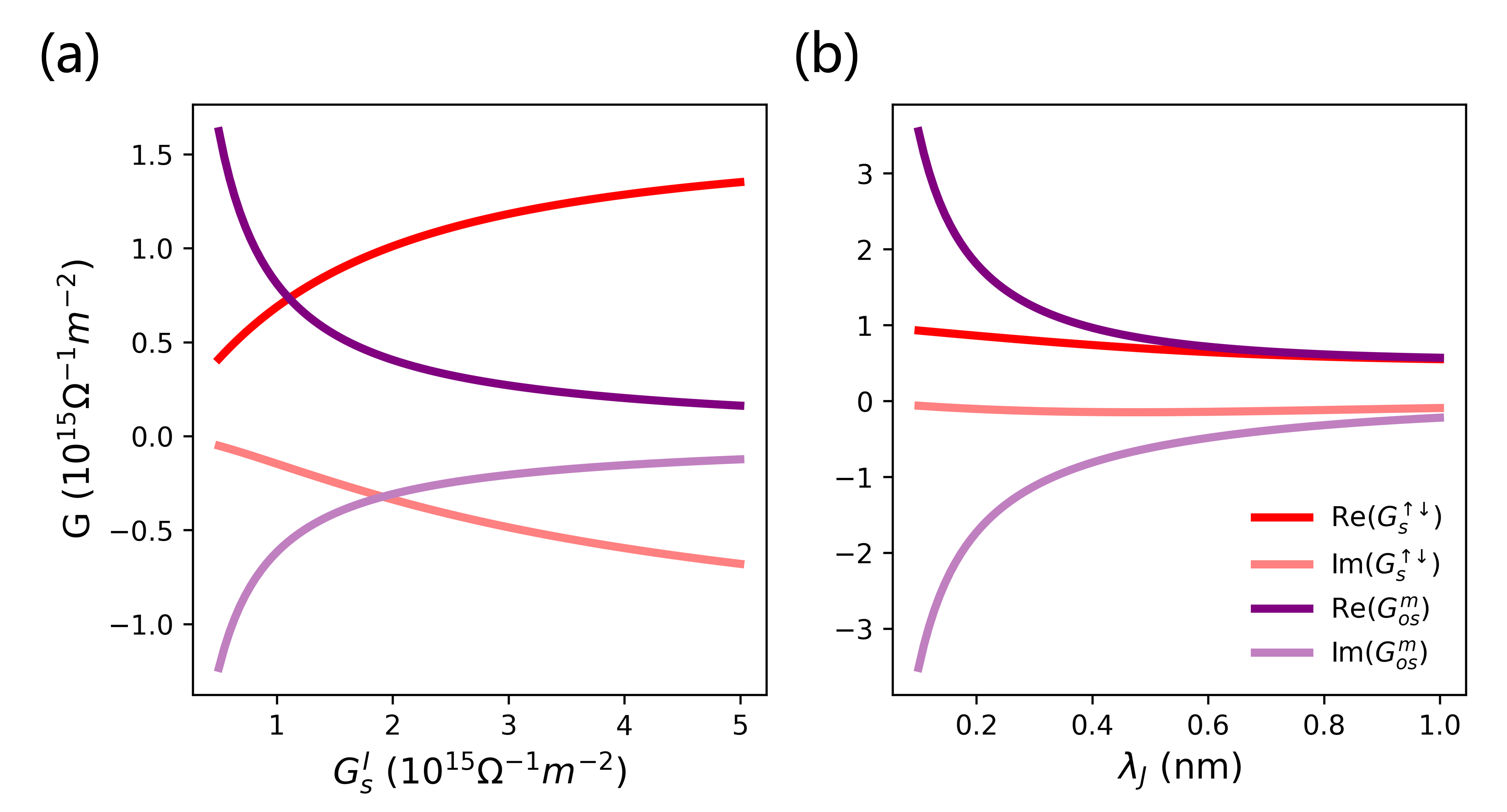}
\caption{(Color online) Spin-mixing conductance (red), $G_{\uparrow\downarrow}$, and orbit-spin-mixing conductance (purple), $G_{os}^m$, as a function of (a) the interfacial spin conductance and (b) the spin precession length. Without loss of generality, we fix $\lambda_J=0.5nm$ in (a) and $G_s=1\times10^{15}\Omega^{-1}m^{-2}$ in (b). The other parameters are set to  $G_{os}=5\times10^{14}\Omega^{-1}m^{-2}$, $\sigma_s^{\rm F}=1\times10^6\Omega^{-1}m^{-1}$, $\lambda_{sf}=3nm$, and $\lambda_{\phi}=1nm$.} 
\label{conductance} 
\end{figure}


This concept enables us to propose a phenomenological expression for the orbital pumping, recently investigated using microscopic models \cite{Han2023,Go2023,Pezo2024} and studied experimentally \cite{Hayashi2024}. Based on Eqs. \eqref{interface_orbital_boundary4}-\eqref{interface_orbital_boundary5}, we obtain the response tensor that connects generalized currents with generalized forces \cite{Brataas2012b}
\begin{eqnarray}
\left(\begin{matrix}
{\bf I}_s\\
{\bf I}_o\\
\partial_t{\bf m}
\end{matrix}\right)=\left(\begin{matrix}
\frac{A}{e^2}G_s^{\uparrow\downarrow} & \frac{A}{e^2}G^m_{os} &{\cal L}_{ms}\\
\frac{A}{e^2}G^m_{so} & \frac{A}{e^2}G_{o} &{\cal L}_{mo}\\
{\cal L}_{sm} & {\cal L}_{om} &\frac{\gamma}{M_s\Omega}{\bf m}\times
\end{matrix}\right)\left(\begin{matrix}
e\Delta{\bm\mu}_s\\
e\Delta{\bm\mu}_o\\
{\bf F}_{\bf m}
\end{matrix}\right)\label{eq:Onsager}
\end{eqnarray}
Here, ${\bf I}_{s,o}=(A/e){\bf J}_{s,o}$ is particle current associated with the spin and orbital current density, $A$ is the contact area between N and F, and $G^m_{os}$ and $G^m_{so}$ are the effective orbit-spin- and spin-orbit-mixing conductances introduced above. ${\bf F}_{\bf m}=\partial_{\bf m}W\approx-(M_s\Omega/\gamma){\bf m}\times\partial_t{\bf m}$ is the magnetic driving force, where $M_s$ is the saturation magnetization (in A/m) and $\Omega$ is the volume of the magnetic layer. The matrix elements are $3\times3$ tensors in angular momentum space. The torque is defined ${\bf T}=(\mu_B/edM_s){\bf m}\times({\bf J}_s\times{\bf m})$ and is readily connected to $e\Delta{\bm\mu}_{s,o}$ via Eq. \eqref{eq:Onsager}, giving
\begin{eqnarray}
{\cal L}_{sm}=\frac{\mu_B}{e^2dM_s}G_s^{\uparrow\downarrow},\;{\cal L}_{om}=\frac{\mu_B}{e^2dM_s}G^m_{os}.
\end{eqnarray}
Applying Onsager's reciprocity \cite{Brataas2012b}, we obtain ${\cal L}_{ms}={\cal L}_{sm}$ and ${\cal L}_{mo}=(\mu_B/e^2dM_s)G^m_{so}$. The first term produces the well-known spin pumping, ${\bf J}_s=(\hbar/e) G^{\uparrow\downarrow}_{s}{\bf m}\times\partial_t{\bf m}$ whereas the second one quantifies the orbital pumping, i.e., ${\bf J}_o=(\hbar/e)G^m_{so}{\bf m}\times\partial_t{\bf m}$.

\section{CONCLUSION}

In this work, we have developed a phenomenological model to describe the orbital torque in magnetic bilayers accounting for spin-orbit interconversion in F and at the N/F interface. It enables us to clarify the parameters that control the orbital torque, demonstrating the interplay between orbital current injection and spin current backflow. The current and potential profiles indicate that, despite its short dephasing length, the spin current exhibits a long propagation in F, driven by the orbit-spin polarization of the smoothly diffusing orbital current. Our calculations also reveal that a long orbital relaxation length in F results in a smoother thickness dependence but also in a larger saturation. This suggests that the orbital torque scales with the orbital relaxation length. Finally, we discuss the concept of orbital-mixing conductance and introduce the orbit-spin- and spin-orbit-mixing conductances that govern the orbital torque and orbital pumping, respectively.

\begin{acknowledgments}
X.N. was supported by the China Scholarship Council Program. W.Z. was supported by the National Key Research and Development Program of China (Grants No. 2022YFB4400200),
National Natural Science Foundation of China (Grant No.
92164206, No. 52261145694, and No. 52121001), and the
New Cornerstone Science Foundation through the XPLORER
PRIZE. A. M. and H.J. were supported by France 2030
government investment plan managed by the French National Research Agency under grant reference PEPR
SPIN – [SPINTHEORY] ANR-22-EXSP-0009, by
the EIC Pathfinder OPEN grant 101129641 “OBELIX” and by grant ANR-20-CE30-0022-01 “ORION” of the French Agence Nationale de la Recherche.
\end{acknowledgments}

%
%

\bibliography{orbital_mixing}

\begin{thebibliography}{49}%
\makeatletter
\providecommand \@ifxundefined [1]{%
 \@ifx{#1\undefined}
}%
\providecommand \@ifnum [1]{%
 \ifnum #1\expandafter \@firstoftwo
 \else \expandafter \@secondoftwo
 \fi
}%
\providecommand \@ifx [1]{%
 \ifx #1\expandafter \@firstoftwo
 \else \expandafter \@secondoftwo
 \fi
}%
\providecommand \natexlab [1]{#1}%
\providecommand \enquote  [1]{``#1''}%
\providecommand \bibnamefont  [1]{#1}%
\providecommand \bibfnamefont [1]{#1}%
\providecommand \citenamefont [1]{#1}%
\providecommand \href@noop [0]{\@secondoftwo}%
\providecommand \href [0]{\begingroup \@sanitize@url \@href}%
\providecommand \@href[1]{\@@startlink{#1}\@@href}%
\providecommand \@@href[1]{\endgroup#1\@@endlink}%
\providecommand \@sanitize@url [0]{\catcode `\\12\catcode `\$12\catcode
  `\&12\catcode `\#12\catcode `\^12\catcode `\_12\catcode `\%12\relax}%
\providecommand \@@startlink[1]{}%
\providecommand \@@endlink[0]{}%
\providecommand \url  [0]{\begingroup\@sanitize@url \@url }%
\providecommand \@url [1]{\endgroup\@href {#1}{\urlprefix }}%
\providecommand \urlprefix  [0]{URL }%
\providecommand \Eprint [0]{\href }%
\providecommand \doibase [0]{https://doi.org/}%
\providecommand \selectlanguage [0]{\@gobble}%
\providecommand \bibinfo  [0]{\@secondoftwo}%
\providecommand \bibfield  [0]{\@secondoftwo}%
\providecommand \translation [1]{[#1]}%
\providecommand \BibitemOpen [0]{}%
\providecommand \bibitemStop [0]{}%
\providecommand \bibitemNoStop [0]{.\EOS\space}%
\providecommand \EOS [0]{\spacefactor3000\relax}%
\providecommand \BibitemShut  [1]{\csname bibitem#1\endcsname}%
\let\auto@bib@innerbib\@empty
\bibitem [{\citenamefont {Gambardella}\ and\ \citenamefont
  {Miron}(2011)}]{Gambardella2011}%
  \BibitemOpen
  \bibfield  {author} {\bibinfo {author} {\bibfnamefont {P.}~\bibnamefont
  {Gambardella}}\ and\ \bibinfo {author} {\bibfnamefont {I.~M.}\ \bibnamefont
  {Miron}},\ }\bibfield  {title} {\bibinfo {title} {Current-induced spin--orbit
  torques},\ }\href {https://doi.org/10.1098/rsta.2010.0336} {\bibfield
  {journal} {\bibinfo  {journal} {Philosophical Transactions of the Royal
  Society A: Mathematical, Physical and Engineering Sciences}\ }\textbf
  {\bibinfo {volume} {369}},\ \bibinfo {pages} {3175} (\bibinfo {year}
  {2011})}\BibitemShut {NoStop}%
\bibitem [{\citenamefont {Manchon}\ \emph {et~al.}(2019)\citenamefont
  {Manchon}, \citenamefont {{\v Z}elezn{\'y}}, \citenamefont {Miron},
  \citenamefont {Jungwirth}, \citenamefont {Sinova}, \citenamefont {Thiaville},
  \citenamefont {Garello},\ and\ \citenamefont {Gambardella}}]{Manchon2019}%
  \BibitemOpen
  \bibfield  {author} {\bibinfo {author} {\bibfnamefont {A.}~\bibnamefont
  {Manchon}}, \bibinfo {author} {\bibfnamefont {J.}~\bibnamefont {{\v
  Z}elezn{\'y}}}, \bibinfo {author} {\bibfnamefont {I.~M.}\ \bibnamefont
  {Miron}}, \bibinfo {author} {\bibfnamefont {T.}~\bibnamefont {Jungwirth}},
  \bibinfo {author} {\bibfnamefont {J.}~\bibnamefont {Sinova}}, \bibinfo
  {author} {\bibfnamefont {A.}~\bibnamefont {Thiaville}}, \bibinfo {author}
  {\bibfnamefont {K.}~\bibnamefont {Garello}},\ and\ \bibinfo {author}
  {\bibfnamefont {P.}~\bibnamefont {Gambardella}},\ }\bibfield  {title}
  {\bibinfo {title} {Current-induced spin-orbit torques in ferromagnetic and
  antiferromagnetic systems},\ }\href
  {https://doi.org/10.1103/RevModPhys.91.035004} {\bibfield  {journal}
  {\bibinfo  {journal} {Reviews of Modern Physics}\ }\textbf {\bibinfo {volume}
  {91}},\ \bibinfo {pages} {035004} (\bibinfo {year} {2019})}\BibitemShut
  {NoStop}%
\bibitem [{\citenamefont {Dieny}\ \emph {et~al.}(2020)\citenamefont {Dieny},
  \citenamefont {Prejbeanu}, \citenamefont {Garello}, \citenamefont
  {Gambardella}, \citenamefont {Freitas}, \citenamefont {Lehndorff},
  \citenamefont {Raberg}, \citenamefont {Ebels}, \citenamefont {Demokritov},
  \citenamefont {Akerman}, \citenamefont {Deac}, \citenamefont {Pirro},
  \citenamefont {Adelmann}, \citenamefont {Anane}, \citenamefont {Chumak},
  \citenamefont {Hirohata}, \citenamefont {Mangin}, \citenamefont {Valenzuela},
  \citenamefont {Onba{\c s}l{\i}}, \citenamefont {{d'Aquino}}, \citenamefont
  {Prenat}, \citenamefont {Finocchio}, \citenamefont {{Lopez-Diaz}},
  \citenamefont {Chantrell}, \citenamefont {{Chubykalo-Fesenko}},\ and\
  \citenamefont {Bortolotti}}]{Dieny2020}%
  \BibitemOpen
  \bibfield  {author} {\bibinfo {author} {\bibfnamefont {B.}~\bibnamefont
  {Dieny}}, \bibinfo {author} {\bibfnamefont {I.~L.}\ \bibnamefont
  {Prejbeanu}}, \bibinfo {author} {\bibfnamefont {K.}~\bibnamefont {Garello}},
  \bibinfo {author} {\bibfnamefont {P.}~\bibnamefont {Gambardella}}, \bibinfo
  {author} {\bibfnamefont {P.}~\bibnamefont {Freitas}}, \bibinfo {author}
  {\bibfnamefont {R.}~\bibnamefont {Lehndorff}}, \bibinfo {author}
  {\bibfnamefont {W.}~\bibnamefont {Raberg}}, \bibinfo {author} {\bibfnamefont
  {U.}~\bibnamefont {Ebels}}, \bibinfo {author} {\bibfnamefont {S.~O.}\
  \bibnamefont {Demokritov}}, \bibinfo {author} {\bibfnamefont
  {J.}~\bibnamefont {Akerman}}, \bibinfo {author} {\bibfnamefont
  {A.}~\bibnamefont {Deac}}, \bibinfo {author} {\bibfnamefont {P.}~\bibnamefont
  {Pirro}}, \bibinfo {author} {\bibfnamefont {C.}~\bibnamefont {Adelmann}},
  \bibinfo {author} {\bibfnamefont {A.}~\bibnamefont {Anane}}, \bibinfo
  {author} {\bibfnamefont {A.~V.}\ \bibnamefont {Chumak}}, \bibinfo {author}
  {\bibfnamefont {A.}~\bibnamefont {Hirohata}}, \bibinfo {author}
  {\bibfnamefont {S.}~\bibnamefont {Mangin}}, \bibinfo {author} {\bibfnamefont
  {S.~O.}\ \bibnamefont {Valenzuela}}, \bibinfo {author} {\bibfnamefont
  {M.~C.}\ \bibnamefont {Onba{\c s}l{\i}}}, \bibinfo {author} {\bibfnamefont
  {M.}~\bibnamefont {{d'Aquino}}}, \bibinfo {author} {\bibfnamefont
  {G.}~\bibnamefont {Prenat}}, \bibinfo {author} {\bibfnamefont
  {G.}~\bibnamefont {Finocchio}}, \bibinfo {author} {\bibfnamefont
  {L.}~\bibnamefont {{Lopez-Diaz}}}, \bibinfo {author} {\bibfnamefont
  {R.}~\bibnamefont {Chantrell}}, \bibinfo {author} {\bibfnamefont
  {O.}~\bibnamefont {{Chubykalo-Fesenko}}},\ and\ \bibinfo {author}
  {\bibfnamefont {P.}~\bibnamefont {Bortolotti}},\ }\bibfield  {title}
  {\bibinfo {title} {Opportunities and challenges for spintronics in the
  microelectronics industry},\ }\href
  {https://doi.org/10.1038/s41928-020-0461-5} {\bibfield  {journal} {\bibinfo
  {journal} {Nature Electronics}\ }\textbf {\bibinfo {volume} {3}},\ \bibinfo
  {pages} {446} (\bibinfo {year} {2020})}\BibitemShut {NoStop}%
\bibitem [{\citenamefont {Guo}\ \emph {et~al.}(2021)\citenamefont {Guo},
  \citenamefont {Yin}, \citenamefont {Bai}, \citenamefont {Zhu}, \citenamefont
  {Shi}, \citenamefont {Wang}, \citenamefont {Cao},\ and\ \citenamefont
  {Zhao}}]{Guo2021}%
  \BibitemOpen
  \bibfield  {author} {\bibinfo {author} {\bibfnamefont {Z.}~\bibnamefont
  {Guo}}, \bibinfo {author} {\bibfnamefont {J.}~\bibnamefont {Yin}}, \bibinfo
  {author} {\bibfnamefont {Y.}~\bibnamefont {Bai}}, \bibinfo {author}
  {\bibfnamefont {D.}~\bibnamefont {Zhu}}, \bibinfo {author} {\bibfnamefont
  {K.}~\bibnamefont {Shi}}, \bibinfo {author} {\bibfnamefont {G.}~\bibnamefont
  {Wang}}, \bibinfo {author} {\bibfnamefont {K.}~\bibnamefont {Cao}},\ and\
  \bibinfo {author} {\bibfnamefont {W.}~\bibnamefont {Zhao}},\ }\bibfield
  {title} {\bibinfo {title} {Spintronics for energy- efficient computing: An
  overview and outlook},\ }\href {https://doi.org/10.1109/JPROC.2021.3084997}
  {\bibfield  {journal} {\bibinfo  {journal} {Proceedings of the IEEE}\
  }\textbf {\bibinfo {volume} {109}},\ \bibinfo {pages} {1398} (\bibinfo {year}
  {2021})}\BibitemShut {NoStop}%
\bibitem [{\citenamefont {Hoffmann}(2013)}]{Hoffmann2013}%
  \BibitemOpen
  \bibfield  {author} {\bibinfo {author} {\bibfnamefont {A.}~\bibnamefont
  {Hoffmann}},\ }\bibfield  {title} {\bibinfo {title} {Spin hall effects in
  metals},\ }\href {https://doi.org/10.1109/TMAG.2013.2262947} {\bibfield
  {journal} {\bibinfo  {journal} {IEEE Transactions on Magnetics}\ }\textbf
  {\bibinfo {volume} {49}},\ \bibinfo {pages} {5172} (\bibinfo {year}
  {2013})}\BibitemShut {NoStop}%
\bibitem [{\citenamefont {Sinova}\ \emph {et~al.}(2015)\citenamefont {Sinova},
  \citenamefont {Valenzuela}, \citenamefont {Wunderlich}, \citenamefont
  {Back},\ and\ \citenamefont {Jungwirth}}]{Sinova2015}%
  \BibitemOpen
  \bibfield  {author} {\bibinfo {author} {\bibfnamefont {J.}~\bibnamefont
  {Sinova}}, \bibinfo {author} {\bibfnamefont {S.~O.}\ \bibnamefont
  {Valenzuela}}, \bibinfo {author} {\bibfnamefont {J.}~\bibnamefont
  {Wunderlich}}, \bibinfo {author} {\bibfnamefont {C.~H.}\ \bibnamefont
  {Back}},\ and\ \bibinfo {author} {\bibfnamefont {T.}~\bibnamefont
  {Jungwirth}},\ }\bibfield  {title} {\bibinfo {title} {Spin hall effects},\
  }\href {https://doi.org/10.1103/RevModPhys.87.1213} {\bibfield  {journal}
  {\bibinfo  {journal} {Reviews of Modern Physics}\ }\textbf {\bibinfo {volume}
  {87}},\ \bibinfo {pages} {1213} (\bibinfo {year} {2015})}\BibitemShut
  {NoStop}%
\bibitem [{\citenamefont {Manchon}\ \emph {et~al.}(2015)\citenamefont
  {Manchon}, \citenamefont {Koo}, \citenamefont {Nitta}, \citenamefont
  {Frolov},\ and\ \citenamefont {Duine}}]{Manchon2015}%
  \BibitemOpen
  \bibfield  {author} {\bibinfo {author} {\bibfnamefont {A.}~\bibnamefont
  {Manchon}}, \bibinfo {author} {\bibfnamefont {H.~C.}\ \bibnamefont {Koo}},
  \bibinfo {author} {\bibfnamefont {J.}~\bibnamefont {Nitta}}, \bibinfo
  {author} {\bibfnamefont {S.~M.}\ \bibnamefont {Frolov}},\ and\ \bibinfo
  {author} {\bibfnamefont {R.~A.}\ \bibnamefont {Duine}},\ }\bibfield  {title}
  {\bibinfo {title} {New perspectives for rashba spin--orbit coupling},\ }\href
  {https://doi.org/10.1038/nmat4360} {\bibfield  {journal} {\bibinfo  {journal}
  {Nature Materials}\ }\textbf {\bibinfo {volume} {14}},\ \bibinfo {pages}
  {871} (\bibinfo {year} {2015})}\BibitemShut {NoStop}%
\bibitem [{\citenamefont {Bihlmayer}\ \emph {et~al.}(2022)\citenamefont
  {Bihlmayer}, \citenamefont {No{\"e}l}, \citenamefont {Vyalikh}, \citenamefont
  {Chulkov},\ and\ \citenamefont {Manchon}}]{Bihlmayer2022}%
  \BibitemOpen
  \bibfield  {author} {\bibinfo {author} {\bibfnamefont {G.}~\bibnamefont
  {Bihlmayer}}, \bibinfo {author} {\bibfnamefont {P.}~\bibnamefont {No{\"e}l}},
  \bibinfo {author} {\bibfnamefont {D.~V.}\ \bibnamefont {Vyalikh}}, \bibinfo
  {author} {\bibfnamefont {E.~V.}\ \bibnamefont {Chulkov}},\ and\ \bibinfo
  {author} {\bibfnamefont {A.}~\bibnamefont {Manchon}},\ }\bibfield  {title}
  {\bibinfo {title} {Rashba-like physics in condensed matter},\ }\href
  {https://doi.org/10.1038/s42254-022-00490-y} {\bibfield  {journal} {\bibinfo
  {journal} {Nature Reviews Physics}\ }\textbf {\bibinfo {volume} {4}},\
  \bibinfo {pages} {642} (\bibinfo {year} {2022})}\BibitemShut {NoStop}%
\bibitem [{\citenamefont {Zhang}\ and\ \citenamefont {Yang}(2005)}]{Zhang2005}%
  \BibitemOpen
  \bibfield  {author} {\bibinfo {author} {\bibfnamefont {S.}~\bibnamefont
  {Zhang}}\ and\ \bibinfo {author} {\bibfnamefont {Z.}~\bibnamefont {Yang}},\
  }\bibfield  {title} {\bibinfo {title} {Intrinsic spin and orbital angular
  momentum hall effect},\ }\href
  {https://doi.org/10.1103/PhysRevLett.94.066602} {\bibfield  {journal}
  {\bibinfo  {journal} {Physical Review Letters}\ }\textbf {\bibinfo {volume}
  {94}},\ \bibinfo {pages} {066602} (\bibinfo {year} {2005})}\BibitemShut
  {NoStop}%
\bibitem [{\citenamefont {Go}\ \emph {et~al.}(2018)\citenamefont {Go},
  \citenamefont {Jo}, \citenamefont {Kim},\ and\ \citenamefont {Lee}}]{Go2018}%
  \BibitemOpen
  \bibfield  {author} {\bibinfo {author} {\bibfnamefont {D.}~\bibnamefont
  {Go}}, \bibinfo {author} {\bibfnamefont {D.}~\bibnamefont {Jo}}, \bibinfo
  {author} {\bibfnamefont {C.}~\bibnamefont {Kim}},\ and\ \bibinfo {author}
  {\bibfnamefont {H.-W.}\ \bibnamefont {Lee}},\ }\bibfield  {title} {\bibinfo
  {title} {Intrinsic spin and orbital hall effects from orbital texture},\
  }\href {https://doi.org/10.1103/PhysRevLett.121.086602} {\bibfield  {journal}
  {\bibinfo  {journal} {Physical Review Letters}\ }\textbf {\bibinfo {volume}
  {121}},\ \bibinfo {pages} {086602} (\bibinfo {year} {2018})}\BibitemShut
  {NoStop}%
\bibitem [{\citenamefont {Go}\ \emph {et~al.}(2017)\citenamefont {Go},
  \citenamefont {Hanke}, \citenamefont {Buhl}, \citenamefont {Freimuth},
  \citenamefont {Bihlmayer}, \citenamefont {Lee}, \citenamefont {Mokrousov},\
  and\ \citenamefont {Bl{\"u}gel}}]{Go2017}%
  \BibitemOpen
  \bibfield  {author} {\bibinfo {author} {\bibfnamefont {D.}~\bibnamefont
  {Go}}, \bibinfo {author} {\bibfnamefont {J.-P.}\ \bibnamefont {Hanke}},
  \bibinfo {author} {\bibfnamefont {P.~M.}\ \bibnamefont {Buhl}}, \bibinfo
  {author} {\bibfnamefont {F.}~\bibnamefont {Freimuth}}, \bibinfo {author}
  {\bibfnamefont {G.}~\bibnamefont {Bihlmayer}}, \bibinfo {author}
  {\bibfnamefont {H.-W.}\ \bibnamefont {Lee}}, \bibinfo {author} {\bibfnamefont
  {Y.}~\bibnamefont {Mokrousov}},\ and\ \bibinfo {author} {\bibfnamefont
  {S.}~\bibnamefont {Bl{\"u}gel}},\ }\bibfield  {title} {\bibinfo {title}
  {Toward surface orbitronics: Giant orbital magnetism from the orbital rashba
  effect at the surface of sp-metals},\ }\href
  {https://doi.org/10.1038/srep46742} {\bibfield  {journal} {\bibinfo
  {journal} {Scientific Reports}\ }\textbf {\bibinfo {volume} {7}},\ \bibinfo
  {pages} {46742} (\bibinfo {year} {2017})}\BibitemShut {NoStop}%
\bibitem [{\citenamefont {Go}\ \emph {et~al.}(2021)\citenamefont {Go},
  \citenamefont {Jo}, \citenamefont {Gao}, \citenamefont {Ando}, \citenamefont
  {Bl{\"u}gel}, \citenamefont {Lee},\ and\ \citenamefont {Mokrousov}}]{Go2021}%
  \BibitemOpen
  \bibfield  {author} {\bibinfo {author} {\bibfnamefont {D.}~\bibnamefont
  {Go}}, \bibinfo {author} {\bibfnamefont {D.}~\bibnamefont {Jo}}, \bibinfo
  {author} {\bibfnamefont {T.}~\bibnamefont {Gao}}, \bibinfo {author}
  {\bibfnamefont {K.}~\bibnamefont {Ando}}, \bibinfo {author} {\bibfnamefont
  {S.}~\bibnamefont {Bl{\"u}gel}}, \bibinfo {author} {\bibfnamefont {H.-W.}\
  \bibnamefont {Lee}},\ and\ \bibinfo {author} {\bibfnamefont {Y.}~\bibnamefont
  {Mokrousov}},\ }\bibfield  {title} {\bibinfo {title} {Orbital rashba effect
  in a surface-oxidized cu film},\ }\href
  {https://doi.org/10.1103/PhysRevB.103.L121113} {\bibfield  {journal}
  {\bibinfo  {journal} {Physical Review B}\ }\textbf {\bibinfo {volume}
  {103}},\ \bibinfo {pages} {L121113} (\bibinfo {year} {2021})}\BibitemShut
  {NoStop}%
\bibitem [{\citenamefont {Go}\ and\ \citenamefont {Lee}(2020)}]{Go2020}%
  \BibitemOpen
  \bibfield  {author} {\bibinfo {author} {\bibfnamefont {D.}~\bibnamefont
  {Go}}\ and\ \bibinfo {author} {\bibfnamefont {H.-W.}\ \bibnamefont {Lee}},\
  }\bibfield  {title} {\bibinfo {title} {Orbital torque: Torque generation by
  orbital current injection},\ }\href
  {https://doi.org/10.1103/PhysRevResearch.2.013177} {\bibfield  {journal}
  {\bibinfo  {journal} {Physical Review Research}\ }\textbf {\bibinfo {volume}
  {2}},\ \bibinfo {pages} {013177} (\bibinfo {year} {2020})}\BibitemShut
  {NoStop}%
\bibitem [{\citenamefont {Ding}\ \emph {et~al.}(2020)\citenamefont {Ding},
  \citenamefont {Ross}, \citenamefont {Go}, \citenamefont {Baldrati},
  \citenamefont {Ren}, \citenamefont {Freimuth}, \citenamefont {Becker},
  \citenamefont {Kammerbauer}, \citenamefont {Yang}, \citenamefont {Jakob},
  \citenamefont {Mokrousov},\ and\ \citenamefont {Kl{\"a}ui}}]{Ding2020}%
  \BibitemOpen
  \bibfield  {author} {\bibinfo {author} {\bibfnamefont {S.}~\bibnamefont
  {Ding}}, \bibinfo {author} {\bibfnamefont {A.}~\bibnamefont {Ross}}, \bibinfo
  {author} {\bibfnamefont {D.}~\bibnamefont {Go}}, \bibinfo {author}
  {\bibfnamefont {L.}~\bibnamefont {Baldrati}}, \bibinfo {author}
  {\bibfnamefont {Z.}~\bibnamefont {Ren}}, \bibinfo {author} {\bibfnamefont
  {F.}~\bibnamefont {Freimuth}}, \bibinfo {author} {\bibfnamefont
  {S.}~\bibnamefont {Becker}}, \bibinfo {author} {\bibfnamefont
  {F.}~\bibnamefont {Kammerbauer}}, \bibinfo {author} {\bibfnamefont
  {J.}~\bibnamefont {Yang}}, \bibinfo {author} {\bibfnamefont {G.}~\bibnamefont
  {Jakob}}, \bibinfo {author} {\bibfnamefont {Y.}~\bibnamefont {Mokrousov}},\
  and\ \bibinfo {author} {\bibfnamefont {M.}~\bibnamefont {Kl{\"a}ui}},\
  }\bibfield  {title} {\bibinfo {title} {Harnessing orbital-to-spin conversion
  of interfacial orbital currents for efficient spin-orbit torques},\ }\href
  {https://doi.org/10.1103/PhysRevLett.125.177201} {\bibfield  {journal}
  {\bibinfo  {journal} {Physical Review Letters}\ }\textbf {\bibinfo {volume}
  {125}},\ \bibinfo {pages} {177201} (\bibinfo {year} {2020})}\BibitemShut
  {NoStop}%
\bibitem [{\citenamefont {Lee}\ \emph {et~al.}(2021{\natexlab{a}})\citenamefont
  {Lee}, \citenamefont {Go}, \citenamefont {Park}, \citenamefont {Jeong},
  \citenamefont {Ko}, \citenamefont {Yun}, \citenamefont {Jo}, \citenamefont
  {Lee}, \citenamefont {Go}, \citenamefont {Oh}, \citenamefont {Kim},
  \citenamefont {Park}, \citenamefont {Min}, \citenamefont {Koo}, \citenamefont
  {Lee}, \citenamefont {Lee},\ and\ \citenamefont {Lee}}]{Lee2021}%
  \BibitemOpen
  \bibfield  {author} {\bibinfo {author} {\bibfnamefont {D.}~\bibnamefont
  {Lee}}, \bibinfo {author} {\bibfnamefont {D.}~\bibnamefont {Go}}, \bibinfo
  {author} {\bibfnamefont {H.-J.}\ \bibnamefont {Park}}, \bibinfo {author}
  {\bibfnamefont {W.}~\bibnamefont {Jeong}}, \bibinfo {author} {\bibfnamefont
  {H.-W.}\ \bibnamefont {Ko}}, \bibinfo {author} {\bibfnamefont
  {D.}~\bibnamefont {Yun}}, \bibinfo {author} {\bibfnamefont {D.}~\bibnamefont
  {Jo}}, \bibinfo {author} {\bibfnamefont {S.}~\bibnamefont {Lee}}, \bibinfo
  {author} {\bibfnamefont {G.}~\bibnamefont {Go}}, \bibinfo {author}
  {\bibfnamefont {J.~H.}\ \bibnamefont {Oh}}, \bibinfo {author} {\bibfnamefont
  {K.-J.}\ \bibnamefont {Kim}}, \bibinfo {author} {\bibfnamefont {B.-G.}\
  \bibnamefont {Park}}, \bibinfo {author} {\bibfnamefont {B.-C.}\ \bibnamefont
  {Min}}, \bibinfo {author} {\bibfnamefont {H.~C.}\ \bibnamefont {Koo}},
  \bibinfo {author} {\bibfnamefont {H.-W.}\ \bibnamefont {Lee}}, \bibinfo
  {author} {\bibfnamefont {O.}~\bibnamefont {Lee}},\ and\ \bibinfo {author}
  {\bibfnamefont {K.-J.}\ \bibnamefont {Lee}},\ }\bibfield  {title} {\bibinfo
  {title} {Orbital torque in magnetic bilayers},\ }\href
  {https://doi.org/10.1038/s41467-021-26650-9} {\bibfield  {journal} {\bibinfo
  {journal} {Nature Communications}\ }\textbf {\bibinfo {volume} {12}},\
  \bibinfo {pages} {6710} (\bibinfo {year} {2021}{\natexlab{a}})}\BibitemShut
  {NoStop}%
\bibitem [{\citenamefont {Lee}\ \emph {et~al.}(2021{\natexlab{b}})\citenamefont
  {Lee}, \citenamefont {Kang}, \citenamefont {Go}, \citenamefont {Kim},
  \citenamefont {Kang}, \citenamefont {Lee}, \citenamefont {Lee}, \citenamefont
  {Kang}, \citenamefont {Lee}, \citenamefont {Mokrousov}, \citenamefont {Kim},
  \citenamefont {Kim}, \citenamefont {Lee},\ and\ \citenamefont
  {Park}}]{Lee2021a}%
  \BibitemOpen
  \bibfield  {author} {\bibinfo {author} {\bibfnamefont {S.}~\bibnamefont
  {Lee}}, \bibinfo {author} {\bibfnamefont {M.-G.}\ \bibnamefont {Kang}},
  \bibinfo {author} {\bibfnamefont {D.}~\bibnamefont {Go}}, \bibinfo {author}
  {\bibfnamefont {D.}~\bibnamefont {Kim}}, \bibinfo {author} {\bibfnamefont
  {J.-H.}\ \bibnamefont {Kang}}, \bibinfo {author} {\bibfnamefont
  {T.}~\bibnamefont {Lee}}, \bibinfo {author} {\bibfnamefont {G.-H.}\
  \bibnamefont {Lee}}, \bibinfo {author} {\bibfnamefont {J.}~\bibnamefont
  {Kang}}, \bibinfo {author} {\bibfnamefont {N.~J.}\ \bibnamefont {Lee}},
  \bibinfo {author} {\bibfnamefont {Y.}~\bibnamefont {Mokrousov}}, \bibinfo
  {author} {\bibfnamefont {S.}~\bibnamefont {Kim}}, \bibinfo {author}
  {\bibfnamefont {K.-J.}\ \bibnamefont {Kim}}, \bibinfo {author} {\bibfnamefont
  {K.-J.}\ \bibnamefont {Lee}},\ and\ \bibinfo {author} {\bibfnamefont {B.-G.}\
  \bibnamefont {Park}},\ }\bibfield  {title} {\bibinfo {title} {Efficient
  conversion of orbital hall current to spin current for spin-orbit torque
  switching},\ }\href {https://doi.org/10.1038/s42005-021-00737-7} {\bibfield
  {journal} {\bibinfo  {journal} {Communications Physics}\ }\textbf {\bibinfo
  {volume} {4}},\ \bibinfo {pages} {234} (\bibinfo {year}
  {2021}{\natexlab{b}})}\BibitemShut {NoStop}%
\bibitem [{\citenamefont {Krishnia}\ \emph {et~al.}(2023)\citenamefont
  {Krishnia}, \citenamefont {Sassi}, \citenamefont {Ajejas}, \citenamefont
  {Sebe}, \citenamefont {Reyren}, \citenamefont {Collin}, \citenamefont
  {Denneulin}, \citenamefont {Kovács}, \citenamefont {Dunin-Borkowski},
  \citenamefont {Fert}, \citenamefont {George}, \citenamefont {Cros},\ and\
  \citenamefont {Jaffrès}}]{Krishnia2023}%
  \BibitemOpen
  \bibfield  {author} {\bibinfo {author} {\bibfnamefont {S.}~\bibnamefont
  {Krishnia}}, \bibinfo {author} {\bibfnamefont {Y.}~\bibnamefont {Sassi}},
  \bibinfo {author} {\bibfnamefont {F.}~\bibnamefont {Ajejas}}, \bibinfo
  {author} {\bibfnamefont {N.}~\bibnamefont {Sebe}}, \bibinfo {author}
  {\bibfnamefont {N.}~\bibnamefont {Reyren}}, \bibinfo {author} {\bibfnamefont
  {S.}~\bibnamefont {Collin}}, \bibinfo {author} {\bibfnamefont
  {T.}~\bibnamefont {Denneulin}}, \bibinfo {author} {\bibfnamefont
  {A.}~\bibnamefont {Kovács}}, \bibinfo {author} {\bibfnamefont {R.~E.}\
  \bibnamefont {Dunin-Borkowski}}, \bibinfo {author} {\bibfnamefont
  {A.}~\bibnamefont {Fert}}, \bibinfo {author} {\bibfnamefont {J.-M.}\
  \bibnamefont {George}}, \bibinfo {author} {\bibfnamefont {V.}~\bibnamefont
  {Cros}},\ and\ \bibinfo {author} {\bibfnamefont {H.}~\bibnamefont
  {Jaffrès}},\ }\bibfield  {title} {\bibinfo {title} {Large interfacial rashba
  interaction generating strong spin-orbit torques in atomically thin metallic
  heterostructures},\ }\href {https://doi.org/10.1021/acs.nanolett.2c05091}
  {\bibfield  {journal} {\bibinfo  {journal} {Nano Letters}\ }\textbf {\bibinfo
  {volume} {23}},\ \bibinfo {pages} {6785} (\bibinfo {year}
  {2023})}\BibitemShut {NoStop}%
\bibitem [{\citenamefont {Yang}\ \emph {et~al.}(2024)\citenamefont {Yang},
  \citenamefont {Xie}, \citenamefont {Zhao}, \citenamefont {Lei}, \citenamefont
  {Zhao},\ and\ \citenamefont {Wei}}]{Yang2024}%
  \BibitemOpen
  \bibfield  {author} {\bibinfo {author} {\bibfnamefont {Y.}~\bibnamefont
  {Yang}}, \bibinfo {author} {\bibfnamefont {Z.}~\bibnamefont {Xie}}, \bibinfo
  {author} {\bibfnamefont {Z.}~\bibnamefont {Zhao}}, \bibinfo {author}
  {\bibfnamefont {N.}~\bibnamefont {Lei}}, \bibinfo {author} {\bibfnamefont
  {J.}~\bibnamefont {Zhao}},\ and\ \bibinfo {author} {\bibfnamefont
  {D.}~\bibnamefont {Wei}},\ }\bibfield  {title} {\bibinfo {title} {Harnessing
  synergy of spin and orbital currents in heavy metal/ferromagnet
  multilayers},\ }\href {https://doi.org/10.1038/s42005-024-01829-w} {\bibfield
   {journal} {\bibinfo  {journal} {Communications Physics}\ }\textbf {\bibinfo
  {volume} {7}},\ \bibinfo {pages} {336} (\bibinfo {year} {2024})}\BibitemShut
  {NoStop}%
\bibitem [{\citenamefont {Go}\ \emph {et~al.}(2023)\citenamefont {Go},
  \citenamefont {Jo}, \citenamefont {Kim}, \citenamefont {Lee}, \citenamefont
  {Kang}, \citenamefont {Park}, \citenamefont {Bl{\"u}gel}, \citenamefont
  {Lee},\ and\ \citenamefont {Mokrousov}}]{Go2023}%
  \BibitemOpen
  \bibfield  {author} {\bibinfo {author} {\bibfnamefont {D.}~\bibnamefont
  {Go}}, \bibinfo {author} {\bibfnamefont {D.}~\bibnamefont {Jo}}, \bibinfo
  {author} {\bibfnamefont {K.-W.}\ \bibnamefont {Kim}}, \bibinfo {author}
  {\bibfnamefont {S.}~\bibnamefont {Lee}}, \bibinfo {author} {\bibfnamefont
  {M.-G.}\ \bibnamefont {Kang}}, \bibinfo {author} {\bibfnamefont {B.-G.}\
  \bibnamefont {Park}}, \bibinfo {author} {\bibfnamefont {S.}~\bibnamefont
  {Bl{\"u}gel}}, \bibinfo {author} {\bibfnamefont {H.-W.}\ \bibnamefont
  {Lee}},\ and\ \bibinfo {author} {\bibfnamefont {Y.}~\bibnamefont
  {Mokrousov}},\ }\bibfield  {title} {\bibinfo {title} {Long-range orbital
  torque by momentum-space hotspots},\ }\href
  {https://doi.org/10.1103/PhysRevLett.130.246701} {\bibfield  {journal}
  {\bibinfo  {journal} {Physical Review Letters}\ }\textbf {\bibinfo {volume}
  {130}},\ \bibinfo {pages} {246701} (\bibinfo {year} {2023})}\BibitemShut
  {NoStop}%
\bibitem [{\citenamefont {Ding}\ \emph {et~al.}(2022)\citenamefont {Ding},
  \citenamefont {No\"el}, \citenamefont {Krishnaswamy},\ and\ \citenamefont
  {Gambardella}}]{Ding2022b}%
  \BibitemOpen
  \bibfield  {author} {\bibinfo {author} {\bibfnamefont {S.}~\bibnamefont
  {Ding}}, \bibinfo {author} {\bibfnamefont {P.}~\bibnamefont {No\"el}},
  \bibinfo {author} {\bibfnamefont {G.~K.}\ \bibnamefont {Krishnaswamy}},\ and\
  \bibinfo {author} {\bibfnamefont {P.}~\bibnamefont {Gambardella}},\
  }\bibfield  {title} {\bibinfo {title} {Unidirectional orbital
  magnetoresistance in light-metal--ferromagnet bilayers},\ }\href
  {https://doi.org/10.1103/PhysRevResearch.4.L032041} {\bibfield  {journal}
  {\bibinfo  {journal} {Phys. Rev. Res.}\ }\textbf {\bibinfo {volume} {4}},\
  \bibinfo {pages} {L032041} (\bibinfo {year} {2022})}\BibitemShut {NoStop}%
\bibitem [{\citenamefont {Ding}\ \emph {et~al.}(2024)\citenamefont {Ding},
  \citenamefont {Wang}, \citenamefont {Legrand}, \citenamefont {Noël},\ and\
  \citenamefont {Gambardella}}]{Ding2024}%
  \BibitemOpen
  \bibfield  {author} {\bibinfo {author} {\bibfnamefont {S.}~\bibnamefont
  {Ding}}, \bibinfo {author} {\bibfnamefont {H.}~\bibnamefont {Wang}}, \bibinfo
  {author} {\bibfnamefont {W.}~\bibnamefont {Legrand}}, \bibinfo {author}
  {\bibfnamefont {P.}~\bibnamefont {Noël}},\ and\ \bibinfo {author}
  {\bibfnamefont {P.}~\bibnamefont {Gambardella}},\ }\bibfield  {title}
  {\bibinfo {title} {Mitigation of gilbert damping in the cofe/cuox orbital
  torque system},\ }\href {https://doi.org/10.1021/acs.nanolett.4c02613}
  {\bibfield  {journal} {\bibinfo  {journal} {Nano Letters}\ }\textbf {\bibinfo
  {volume} {24}},\ \bibinfo {pages} {10251} (\bibinfo {year} {2024})},\
  \bibinfo {note} {pMID: 39133560},\ \Eprint
  {https://arxiv.org/abs/https://doi.org/10.1021/acs.nanolett.4c02613}
  {https://doi.org/10.1021/acs.nanolett.4c02613} \BibitemShut {NoStop}%
\bibitem [{\citenamefont {Hayashi}\ \emph {et~al.}(2023)\citenamefont
  {Hayashi}, \citenamefont {Jo}, \citenamefont {Go}, \citenamefont {Gao},
  \citenamefont {Haku}, \citenamefont {Mokrousov}, \citenamefont {Lee},\ and\
  \citenamefont {Ando}}]{Hayashi2023}%
  \BibitemOpen
  \bibfield  {author} {\bibinfo {author} {\bibfnamefont {H.}~\bibnamefont
  {Hayashi}}, \bibinfo {author} {\bibfnamefont {D.}~\bibnamefont {Jo}},
  \bibinfo {author} {\bibfnamefont {D.}~\bibnamefont {Go}}, \bibinfo {author}
  {\bibfnamefont {T.}~\bibnamefont {Gao}}, \bibinfo {author} {\bibfnamefont
  {S.}~\bibnamefont {Haku}}, \bibinfo {author} {\bibfnamefont {Y.}~\bibnamefont
  {Mokrousov}}, \bibinfo {author} {\bibfnamefont {H.-W.}\ \bibnamefont {Lee}},\
  and\ \bibinfo {author} {\bibfnamefont {K.}~\bibnamefont {Ando}},\ }\bibfield
  {title} {\bibinfo {title} {Observation of long-range orbital transport and
  giant orbital torque},\ }\href {https://doi.org/10.1038/s42005-023-01139-7}
  {\bibfield  {journal} {\bibinfo  {journal} {Communications Physics}\ }\textbf
  {\bibinfo {volume} {6}},\ \bibinfo {pages} {1} (\bibinfo {year}
  {2023})}\BibitemShut {NoStop}%
\bibitem [{\citenamefont {Bose}\ \emph {et~al.}(2023)\citenamefont {Bose},
  \citenamefont {Kammerbauer}, \citenamefont {Gupta}, \citenamefont {Go},
  \citenamefont {Mokrousov}, \citenamefont {Jakob},\ and\ \citenamefont
  {Kl{\"a}ui}}]{Bose2023}%
  \BibitemOpen
  \bibfield  {author} {\bibinfo {author} {\bibfnamefont {A.}~\bibnamefont
  {Bose}}, \bibinfo {author} {\bibfnamefont {F.}~\bibnamefont {Kammerbauer}},
  \bibinfo {author} {\bibfnamefont {R.}~\bibnamefont {Gupta}}, \bibinfo
  {author} {\bibfnamefont {D.}~\bibnamefont {Go}}, \bibinfo {author}
  {\bibfnamefont {Y.}~\bibnamefont {Mokrousov}}, \bibinfo {author}
  {\bibfnamefont {G.}~\bibnamefont {Jakob}},\ and\ \bibinfo {author}
  {\bibfnamefont {M.}~\bibnamefont {Kl{\"a}ui}},\ }\bibfield  {title} {\bibinfo
  {title} {Detection of long-range orbital-hall torques},\ }\href
  {https://doi.org/10.1103/PhysRevB.107.134423} {\bibfield  {journal} {\bibinfo
   {journal} {Physical Review B}\ }\textbf {\bibinfo {volume} {107}},\ \bibinfo
  {pages} {134423} (\bibinfo {year} {2023})}\BibitemShut {NoStop}%
\bibitem [{\citenamefont {Tserkovnyak}\ \emph {et~al.}(2002)\citenamefont
  {Tserkovnyak}, \citenamefont {Brataas},\ and\ \citenamefont
  {Bauer}}]{Tserkovnyak2002}%
  \BibitemOpen
  \bibfield  {author} {\bibinfo {author} {\bibfnamefont {Y.}~\bibnamefont
  {Tserkovnyak}}, \bibinfo {author} {\bibfnamefont {A.}~\bibnamefont
  {Brataas}},\ and\ \bibinfo {author} {\bibfnamefont {G.~E.~W.}\ \bibnamefont
  {Bauer}},\ }\bibfield  {title} {\bibinfo {title} {Spin pumping and
  magnetization dynamics in metallic multilayers},\ }\href
  {https://doi.org/10.1103/PhysRevB.66.224403} {\bibfield  {journal} {\bibinfo
  {journal} {Physical Review B}\ }\textbf {\bibinfo {volume} {66}},\ \bibinfo
  {pages} {224403} (\bibinfo {year} {2002})}\BibitemShut {NoStop}%
\bibitem [{\citenamefont {Haney}\ \emph {et~al.}(2013)\citenamefont {Haney},
  \citenamefont {Lee}, \citenamefont {Lee}, \citenamefont {Manchon},\ and\
  \citenamefont {Stiles}}]{Haney2013}%
  \BibitemOpen
  \bibfield  {author} {\bibinfo {author} {\bibfnamefont {P.~M.}\ \bibnamefont
  {Haney}}, \bibinfo {author} {\bibfnamefont {H.-W.}\ \bibnamefont {Lee}},
  \bibinfo {author} {\bibfnamefont {K.-J.}\ \bibnamefont {Lee}}, \bibinfo
  {author} {\bibfnamefont {A.}~\bibnamefont {Manchon}},\ and\ \bibinfo {author}
  {\bibfnamefont {M.~D.}\ \bibnamefont {Stiles}},\ }\bibfield  {title}
  {\bibinfo {title} {Current induced torques and interfacial spin-orbit
  coupling: Semiclassical modeling},\ }\href
  {https://doi.org/10.1103/PhysRevB.87.174411} {\bibfield  {journal} {\bibinfo
  {journal} {Physical Review B}\ }\textbf {\bibinfo {volume} {87}},\ \bibinfo
  {pages} {174411} (\bibinfo {year} {2013})}\BibitemShut {NoStop}%
\bibitem [{\citenamefont {Tanaka}\ \emph {et~al.}(2008)\citenamefont {Tanaka},
  \citenamefont {Kontani}, \citenamefont {Naito}, \citenamefont {Naito},
  \citenamefont {Hirashima}, \citenamefont {Yamada},\ and\ \citenamefont
  {Inoue}}]{Tanaka2008}%
  \BibitemOpen
  \bibfield  {author} {\bibinfo {author} {\bibfnamefont {T.}~\bibnamefont
  {Tanaka}}, \bibinfo {author} {\bibfnamefont {H.}~\bibnamefont {Kontani}},
  \bibinfo {author} {\bibfnamefont {M.}~\bibnamefont {Naito}}, \bibinfo
  {author} {\bibfnamefont {T.}~\bibnamefont {Naito}}, \bibinfo {author}
  {\bibfnamefont {D.~S.}\ \bibnamefont {Hirashima}}, \bibinfo {author}
  {\bibfnamefont {K.}~\bibnamefont {Yamada}},\ and\ \bibinfo {author}
  {\bibfnamefont {J.}~\bibnamefont {Inoue}},\ }\bibfield  {title} {\bibinfo
  {title} {Intrinsic spin hall effect and orbital hall effect in 4d and 5d
  transition metals},\ }\href {https://doi.org/10.1103/PhysRevB.77.165117}
  {\bibfield  {journal} {\bibinfo  {journal} {Physical Review B}\ }\textbf
  {\bibinfo {volume} {77}},\ \bibinfo {pages} {165117} (\bibinfo {year}
  {2008})}\BibitemShut {NoStop}%
\bibitem [{\citenamefont {Kontani}\ \emph {et~al.}(2009)\citenamefont
  {Kontani}, \citenamefont {Tanaka}, \citenamefont {Hirashima}, \citenamefont
  {Yamada},\ and\ \citenamefont {Inoue}}]{Kontani2009}%
  \BibitemOpen
  \bibfield  {author} {\bibinfo {author} {\bibfnamefont {H.}~\bibnamefont
  {Kontani}}, \bibinfo {author} {\bibfnamefont {T.}~\bibnamefont {Tanaka}},
  \bibinfo {author} {\bibfnamefont {D.~S.}\ \bibnamefont {Hirashima}}, \bibinfo
  {author} {\bibfnamefont {K.}~\bibnamefont {Yamada}},\ and\ \bibinfo {author}
  {\bibfnamefont {J.}~\bibnamefont {Inoue}},\ }\bibfield  {title} {\bibinfo
  {title} {Giant orbital hall effect in transition metals: Origin of large spin
  and anomalous hall effects},\ }\href
  {https://doi.org/10.1103/PhysRevLett.102.016601} {\bibfield  {journal}
  {\bibinfo  {journal} {Physical Review Letters}\ }\textbf {\bibinfo {volume}
  {102}},\ \bibinfo {pages} {016601} (\bibinfo {year} {2009})}\BibitemShut
  {NoStop}%
\bibitem [{\citenamefont {Salemi}\ and\ \citenamefont
  {Oppeneer}(2022)}]{Salemi2022}%
  \BibitemOpen
  \bibfield  {author} {\bibinfo {author} {\bibfnamefont {L.}~\bibnamefont
  {Salemi}}\ and\ \bibinfo {author} {\bibfnamefont {P.~M.}\ \bibnamefont
  {Oppeneer}},\ }\bibfield  {title} {\bibinfo {title} {First-principles theory
  of intrinsic spin and orbital hall and nernst effects in metallic monoatomic
  crystals},\ }\href {https://doi.org/10.1103/PhysRevMaterials.6.095001}
  {\bibfield  {journal} {\bibinfo  {journal} {Physical Review Materials}\
  }\textbf {\bibinfo {volume} {6}},\ \bibinfo {pages} {095001} (\bibinfo {year}
  {2022})}\BibitemShut {NoStop}%
\bibitem [{\citenamefont {Pezo}\ \emph {et~al.}(2022)\citenamefont {Pezo},
  \citenamefont {Garc{\'i}a~Ovalle},\ and\ \citenamefont {Manchon}}]{Pezo2022}%
  \BibitemOpen
  \bibfield  {author} {\bibinfo {author} {\bibfnamefont {A.}~\bibnamefont
  {Pezo}}, \bibinfo {author} {\bibfnamefont {D.}~\bibnamefont
  {Garc{\'i}a~Ovalle}},\ and\ \bibinfo {author} {\bibfnamefont
  {A.}~\bibnamefont {Manchon}},\ }\bibfield  {title} {\bibinfo {title} {Orbital
  hall effect in crystals: Interatomic versus intra-atomic contributions},\
  }\href {https://doi.org/10.1103/PhysRevB.106.104414} {\bibfield  {journal}
  {\bibinfo  {journal} {Physical Review B}\ }\textbf {\bibinfo {volume}
  {106}},\ \bibinfo {pages} {104414} (\bibinfo {year} {2022})}\BibitemShut
  {NoStop}%
\bibitem [{\citenamefont {Go}\ \emph {et~al.}(2024)\citenamefont {Go},
  \citenamefont {Lee}, \citenamefont {Oppeneer}, \citenamefont {Bl{\"u}gel},\
  and\ \citenamefont {Mokrousov}}]{Go2024}%
  \BibitemOpen
  \bibfield  {author} {\bibinfo {author} {\bibfnamefont {D.}~\bibnamefont
  {Go}}, \bibinfo {author} {\bibfnamefont {H.-W.}\ \bibnamefont {Lee}},
  \bibinfo {author} {\bibfnamefont {P.~M.}\ \bibnamefont {Oppeneer}}, \bibinfo
  {author} {\bibfnamefont {S.}~\bibnamefont {Bl{\"u}gel}},\ and\ \bibinfo
  {author} {\bibfnamefont {Y.}~\bibnamefont {Mokrousov}},\ }\bibfield  {title}
  {\bibinfo {title} {First-principles calculation of orbital hall effect by
  wannier interpolation: Role of orbital dependence of the anomalous
  position},\ }\href {https://doi.org/10.1103/PhysRevB.109.174435} {\bibfield
  {journal} {\bibinfo  {journal} {Physical Review B}\ }\textbf {\bibinfo
  {volume} {109}},\ \bibinfo {pages} {174435} (\bibinfo {year}
  {2024})}\BibitemShut {NoStop}%
\bibitem [{\citenamefont {Sala}\ and\ \citenamefont
  {Gambardella}(2022)}]{Sala2022}%
  \BibitemOpen
  \bibfield  {author} {\bibinfo {author} {\bibfnamefont {G.}~\bibnamefont
  {Sala}}\ and\ \bibinfo {author} {\bibfnamefont {P.}~\bibnamefont
  {Gambardella}},\ }\bibfield  {title} {\bibinfo {title} {Giant orbital hall
  effect and orbital-to-spin conversion in 3d, 5d, and 4f metallic
  heterostructures},\ }\href {https://doi.org/10.1103/PhysRevResearch.4.033037}
  {\bibfield  {journal} {\bibinfo  {journal} {Physical Review Research}\
  }\textbf {\bibinfo {volume} {4}},\ \bibinfo {pages} {033037} (\bibinfo {year}
  {2022})}\BibitemShut {NoStop}%
\bibitem [{\citenamefont {Han}\ \emph {et~al.}(2022)\citenamefont {Han},
  \citenamefont {Lee},\ and\ \citenamefont {Kim}}]{Han2022}%
  \BibitemOpen
  \bibfield  {author} {\bibinfo {author} {\bibfnamefont {S.}~\bibnamefont
  {Han}}, \bibinfo {author} {\bibfnamefont {H.~W.}\ \bibnamefont {Lee}},\ and\
  \bibinfo {author} {\bibfnamefont {K.~W.}\ \bibnamefont {Kim}},\ }\bibfield
  {title} {\bibinfo {title} {Orbital dynamics in centrosymmetric systems},\
  }\bibfield  {journal} {\bibinfo  {journal} {Physical Review Letters}\
  }\textbf {\bibinfo {volume} {128}},\ \href
  {https://doi.org/10.1103/PhysRevLett.128.176601}
  {10.1103/PhysRevLett.128.176601} (\bibinfo {year} {2022})\BibitemShut
  {NoStop}%
\bibitem [{\citenamefont {Ning}\ \emph {et~al.}(2024)\citenamefont {Ning},
  \citenamefont {Pezo}, \citenamefont {Kim}, \citenamefont {Zhao},
  \citenamefont {Lee},\ and\ \citenamefont {Manchon}}]{Ning2024}%
  \BibitemOpen
  \bibfield  {author} {\bibinfo {author} {\bibfnamefont {X.}~\bibnamefont
  {Ning}}, \bibinfo {author} {\bibfnamefont {A.}~\bibnamefont {Pezo}}, \bibinfo
  {author} {\bibfnamefont {K.-W.}\ \bibnamefont {Kim}}, \bibinfo {author}
  {\bibfnamefont {W.}~\bibnamefont {Zhao}}, \bibinfo {author} {\bibfnamefont
  {K.-J.}\ \bibnamefont {Lee}},\ and\ \bibinfo {author} {\bibfnamefont
  {A.}~\bibnamefont {Manchon}},\ }\bibfield  {title} {\bibinfo {title} {Orbital
  diffusion, polarization and swapping in centrosymmetric metals},\ }\href
  {http://arxiv.org/abs/2310.04763} {\bibfield  {journal} {\bibinfo  {journal}
  {arXiv.2310.04763}\ } (\bibinfo {year} {2024})}\BibitemShut {NoStop}%
\bibitem [{\citenamefont {Stiles}\ and\ \citenamefont
  {Zangwill}(2002)}]{Stiles2002}%
  \BibitemOpen
  \bibfield  {author} {\bibinfo {author} {\bibfnamefont {M.}~\bibnamefont
  {Stiles}}\ and\ \bibinfo {author} {\bibfnamefont {A.}~\bibnamefont
  {Zangwill}},\ }\bibfield  {title} {\bibinfo {title} {Anatomy of spin-transfer
  torque},\ }\href {https://doi.org/10.1103/PhysRevB.66.014407} {\bibfield
  {journal} {\bibinfo  {journal} {Physical Review B}\ }\textbf {\bibinfo
  {volume} {66}},\ \bibinfo {pages} {014407} (\bibinfo {year}
  {2002})}\BibitemShut {NoStop}%
\bibitem [{\citenamefont {Zhang}\ \emph {et~al.}(2002)\citenamefont {Zhang},
  \citenamefont {Levy},\ and\ \citenamefont {Fert}}]{Zhang2002}%
  \BibitemOpen
  \bibfield  {author} {\bibinfo {author} {\bibfnamefont {S.}~\bibnamefont
  {Zhang}}, \bibinfo {author} {\bibfnamefont {P.}~\bibnamefont {Levy}},\ and\
  \bibinfo {author} {\bibfnamefont {A.}~\bibnamefont {Fert}},\ }\bibfield
  {title} {\bibinfo {title} {Mechanisms of spin-polarized current-driven
  magnetization switching},\ }\href
  {https://doi.org/10.1103/PhysRevLett.88.236601} {\bibfield  {journal}
  {\bibinfo  {journal} {Physical Review Letters}\ }\textbf {\bibinfo {volume}
  {88}},\ \bibinfo {pages} {236601} (\bibinfo {year} {2002})}\BibitemShut
  {NoStop}%
\bibitem [{\citenamefont {Petitjean}\ \emph {et~al.}(2012)\citenamefont
  {Petitjean}, \citenamefont {Luc},\ and\ \citenamefont
  {Waintal}}]{Petitjean2012}%
  \BibitemOpen
  \bibfield  {author} {\bibinfo {author} {\bibfnamefont {C.}~\bibnamefont
  {Petitjean}}, \bibinfo {author} {\bibfnamefont {D.}~\bibnamefont {Luc}},\
  and\ \bibinfo {author} {\bibfnamefont {X.}~\bibnamefont {Waintal}},\
  }\bibfield  {title} {\bibinfo {title} {Unified drift-diffusion theory for
  transverse spin currents in spin valves, domain walls, and other textured
  magnets},\ }\href {https://doi.org/10.1103/PhysRevLett.109.117204} {\bibfield
   {journal} {\bibinfo  {journal} {Physical Review Letters}\ }\textbf {\bibinfo
  {volume} {109}},\ \bibinfo {pages} {117204} (\bibinfo {year}
  {2012})}\BibitemShut {NoStop}%
\bibitem [{\citenamefont {Lim}\ \emph {et~al.}(2021)\citenamefont {Lim},
  \citenamefont {Khodadadi}, \citenamefont {Li}, \citenamefont {Viehland},
  \citenamefont {Manchon},\ and\ \citenamefont {Emori}}]{Lim2021}%
  \BibitemOpen
  \bibfield  {author} {\bibinfo {author} {\bibfnamefont {Y.}~\bibnamefont
  {Lim}}, \bibinfo {author} {\bibfnamefont {B.}~\bibnamefont {Khodadadi}},
  \bibinfo {author} {\bibfnamefont {J.-F.}\ \bibnamefont {Li}}, \bibinfo
  {author} {\bibfnamefont {D.}~\bibnamefont {Viehland}}, \bibinfo {author}
  {\bibfnamefont {A.}~\bibnamefont {Manchon}},\ and\ \bibinfo {author}
  {\bibfnamefont {S.}~\bibnamefont {Emori}},\ }\bibfield  {title} {\bibinfo
  {title} {Dephasing of transverse spin current in ferrimagnetic alloys},\
  }\href {https://doi.org/10.1103/PhysRevB.103.024443} {\bibfield  {journal}
  {\bibinfo  {journal} {Phys. Rev. B}\ }\textbf {\bibinfo {volume} {103}},\
  \bibinfo {pages} {024443} (\bibinfo {year} {2021})}\BibitemShut {NoStop}%
\bibitem [{\citenamefont {Brataas}\ \emph {et~al.}(2001)\citenamefont
  {Brataas}, \citenamefont {Nazarov},\ and\ \citenamefont
  {Bauer}}]{Brataas2001}%
  \BibitemOpen
  \bibfield  {author} {\bibinfo {author} {\bibfnamefont {A.}~\bibnamefont
  {Brataas}}, \bibinfo {author} {\bibfnamefont {Y.~V.}\ \bibnamefont
  {Nazarov}},\ and\ \bibinfo {author} {\bibfnamefont {G.~E.~W.}\ \bibnamefont
  {Bauer}},\ }\bibfield  {title} {\bibinfo {title} {Spin-transport in
  multi-terminal normal metal-ferromagnet systems with non-collinear
  magnetizations},\ }\href@noop {} {\bibfield  {journal} {\bibinfo  {journal}
  {European Physical Journal B}\ }\textbf {\bibinfo {volume} {110}},\ \bibinfo
  {pages} {99} (\bibinfo {year} {2001})}\BibitemShut {NoStop}%
\bibitem [{\citenamefont {Zwierzycki}\ \emph {et~al.}(2005)\citenamefont
  {Zwierzycki}, \citenamefont {Tserkovnyak}, \citenamefont {Kelly},
  \citenamefont {Brataas},\ and\ \citenamefont {Bauer}}]{Zwierzycki2005}%
  \BibitemOpen
  \bibfield  {author} {\bibinfo {author} {\bibfnamefont {M.}~\bibnamefont
  {Zwierzycki}}, \bibinfo {author} {\bibfnamefont {Y.}~\bibnamefont
  {Tserkovnyak}}, \bibinfo {author} {\bibfnamefont {P.~J.}\ \bibnamefont
  {Kelly}}, \bibinfo {author} {\bibfnamefont {A.}~\bibnamefont {Brataas}},\
  and\ \bibinfo {author} {\bibfnamefont {G.~E.~W.}\ \bibnamefont {Bauer}},\
  }\bibfield  {title} {\bibinfo {title} {First-principles study of
  magnetization relaxation enhancement and spin transfer in thin magnetic
  films},\ }\href {https://doi.org/10.1103/PhysRevB.71.064420} {\bibfield
  {journal} {\bibinfo  {journal} {Physical Review B}\ }\textbf {\bibinfo
  {volume} {71}},\ \bibinfo {pages} {064420} (\bibinfo {year}
  {2005})}\BibitemShut {NoStop}%
\bibitem [{\citenamefont {Brataas}\ \emph {et~al.}(2006)\citenamefont
  {Brataas}, \citenamefont {Bauer},\ and\ \citenamefont {Kelly}}]{Brataas2006}%
  \BibitemOpen
  \bibfield  {author} {\bibinfo {author} {\bibfnamefont {A.}~\bibnamefont
  {Brataas}}, \bibinfo {author} {\bibfnamefont {G.~E.~W.}\ \bibnamefont
  {Bauer}},\ and\ \bibinfo {author} {\bibfnamefont {P.}~\bibnamefont {Kelly}},\
  }\bibfield  {title} {\bibinfo {title} {Non-collinear magnetoelectronics},\
  }\href {https://doi.org/10.1016/j.physrep.2006.01.001} {\bibfield  {journal}
  {\bibinfo  {journal} {Physics Reports}\ }\textbf {\bibinfo {volume} {427}},\
  \bibinfo {pages} {157} (\bibinfo {year} {2006})}\BibitemShut {NoStop}%
\bibitem [{\citenamefont {Shchelushkin}\ and\ \citenamefont
  {Brataas}(2005)}]{Shchelushkin2005}%
  \BibitemOpen
  \bibfield  {author} {\bibinfo {author} {\bibfnamefont {R.}~\bibnamefont
  {Shchelushkin}}\ and\ \bibinfo {author} {\bibfnamefont {A.}~\bibnamefont
  {Brataas}},\ }\bibfield  {title} {\bibinfo {title} {Spin hall effect, hall
  effect, and spin precession in diffusive normal metals},\ }\href
  {https://doi.org/10.1103/PhysRevB.72.073110} {\bibfield  {journal} {\bibinfo
  {journal} {Physical Review B}\ }\textbf {\bibinfo {volume} {72}},\ \bibinfo
  {pages} {073110} (\bibinfo {year} {2005})}\BibitemShut {NoStop}%
\bibitem [{\citenamefont {Pauyac}\ \emph {et~al.}(2018)\citenamefont {Pauyac},
  \citenamefont {Chshiev}, \citenamefont {Manchon},\ and\ \citenamefont
  {Nikolaev}}]{Pauyac2018}%
  \BibitemOpen
  \bibfield  {author} {\bibinfo {author} {\bibfnamefont {C.~O.}\ \bibnamefont
  {Pauyac}}, \bibinfo {author} {\bibfnamefont {M.}~\bibnamefont {Chshiev}},
  \bibinfo {author} {\bibfnamefont {A.}~\bibnamefont {Manchon}},\ and\ \bibinfo
  {author} {\bibfnamefont {S.~A.}\ \bibnamefont {Nikolaev}},\ }\bibfield
  {title} {\bibinfo {title} {Spin hall and spin swapping torques in diffusive
  ferromagnets},\ }\href {https://doi.org/10.1103/PhysRevLett.120.176802}
  {\bibfield  {journal} {\bibinfo  {journal} {Physical Review Letters}\
  }\textbf {\bibinfo {volume} {120}},\ \bibinfo {pages} {176802} (\bibinfo
  {year} {2018})}\BibitemShut {NoStop}%
\bibitem [{\citenamefont {Sohn}\ \emph {et~al.}(2024)\citenamefont {Sohn},
  \citenamefont {Lee},\ and\ \citenamefont {Lee}}]{Sohn2024}%
  \BibitemOpen
  \bibfield  {author} {\bibinfo {author} {\bibfnamefont {J.}~\bibnamefont
  {Sohn}}, \bibinfo {author} {\bibfnamefont {J.~M.}\ \bibnamefont {Lee}},\ and\
  \bibinfo {author} {\bibfnamefont {H.~W.}\ \bibnamefont {Lee}},\ }\bibfield
  {title} {\bibinfo {title} {Dyakonov-perel-like orbital and spin relaxations
  in centrosymmetric systems},\ }\href
  {https://doi.org/10.1103/PhysRevLett.132.246301} {\bibfield  {journal}
  {\bibinfo  {journal} {Physical Review Letters}\ }\textbf {\bibinfo {volume}
  {132}},\ \bibinfo {pages} {246301} (\bibinfo {year} {2024})}\BibitemShut
  {NoStop}%
\bibitem [{\citenamefont {Go}\ \emph {et~al.}(2020)\citenamefont {Go},
  \citenamefont {Freimuth}, \citenamefont {Hanke}, \citenamefont {Xue},
  \citenamefont {Gomonay}, \citenamefont {Lee}, \citenamefont {Bl{\"u}gel},
  \citenamefont {Haney}, \citenamefont {Lee},\ and\ \citenamefont
  {Mokrousov}}]{Go2020a}%
  \BibitemOpen
  \bibfield  {author} {\bibinfo {author} {\bibfnamefont {D.}~\bibnamefont
  {Go}}, \bibinfo {author} {\bibfnamefont {F.}~\bibnamefont {Freimuth}},
  \bibinfo {author} {\bibfnamefont {J.-P.}\ \bibnamefont {Hanke}}, \bibinfo
  {author} {\bibfnamefont {F.}~\bibnamefont {Xue}}, \bibinfo {author}
  {\bibfnamefont {O.}~\bibnamefont {Gomonay}}, \bibinfo {author} {\bibfnamefont
  {K.-J.}\ \bibnamefont {Lee}}, \bibinfo {author} {\bibfnamefont
  {S.}~\bibnamefont {Bl{\"u}gel}}, \bibinfo {author} {\bibfnamefont {P.~M.}\
  \bibnamefont {Haney}}, \bibinfo {author} {\bibfnamefont {H.-W.}\ \bibnamefont
  {Lee}},\ and\ \bibinfo {author} {\bibfnamefont {Y.}~\bibnamefont
  {Mokrousov}},\ }\bibfield  {title} {\bibinfo {title} {Theory of
  current-induced angular momentum transfer dynamics in spin-orbit coupled
  systems},\ }\href {https://doi.org/10.1103/PhysRevResearch.2.033401}
  {\bibfield  {journal} {\bibinfo  {journal} {Physical Review Research}\
  }\textbf {\bibinfo {volume} {2}},\ \bibinfo {pages} {033401} (\bibinfo {year}
  {2020})}\BibitemShut {NoStop}%
\bibitem [{\citenamefont {Krishnia}\ \emph {et~al.}(2024)\citenamefont
  {Krishnia}, \citenamefont {Bony}, \citenamefont {Rongione}, \citenamefont
  {Vicente-Arche}, \citenamefont {Denneulin}, \citenamefont {Pezo},
  \citenamefont {Lu}, \citenamefont {Dunin-Borkowski}, \citenamefont {Collin},
  \citenamefont {Fert}, \citenamefont {George}, \citenamefont {Reyren},
  \citenamefont {Cros},\ and\ \citenamefont {Jaffrès}}]{Krishnia2024}%
  \BibitemOpen
  \bibfield  {author} {\bibinfo {author} {\bibfnamefont {S.}~\bibnamefont
  {Krishnia}}, \bibinfo {author} {\bibfnamefont {B.}~\bibnamefont {Bony}},
  \bibinfo {author} {\bibfnamefont {E.}~\bibnamefont {Rongione}}, \bibinfo
  {author} {\bibfnamefont {L.~M.}\ \bibnamefont {Vicente-Arche}}, \bibinfo
  {author} {\bibfnamefont {T.}~\bibnamefont {Denneulin}}, \bibinfo {author}
  {\bibfnamefont {A.}~\bibnamefont {Pezo}}, \bibinfo {author} {\bibfnamefont
  {Y.}~\bibnamefont {Lu}}, \bibinfo {author} {\bibfnamefont {R.~E.}\
  \bibnamefont {Dunin-Borkowski}}, \bibinfo {author} {\bibfnamefont
  {S.}~\bibnamefont {Collin}}, \bibinfo {author} {\bibfnamefont
  {A.}~\bibnamefont {Fert}}, \bibinfo {author} {\bibfnamefont {J.-M.}\
  \bibnamefont {George}}, \bibinfo {author} {\bibfnamefont {N.}~\bibnamefont
  {Reyren}}, \bibinfo {author} {\bibfnamefont {V.}~\bibnamefont {Cros}},\ and\
  \bibinfo {author} {\bibfnamefont {H.}~\bibnamefont {Jaffrès}},\ }\bibfield
  {title} {\bibinfo {title} {Quantifying the large contribution from orbital
  rashba–edelstein effect to the effective damping-like torque on
  magnetization},\ }\href {https://doi.org/10.1063/5.0198970} {\bibfield
  {journal} {\bibinfo  {journal} {APL Materials}\ }\textbf {\bibinfo {volume}
  {12}},\ \bibinfo {pages} {051105} (\bibinfo {year} {2024})}\BibitemShut
  {NoStop}%
\bibitem [{\citenamefont {Han}\ \emph {et~al.}(2023)\citenamefont {Han},
  \citenamefont {Ko}, \citenamefont {Oh}, \citenamefont {Lee}, \citenamefont
  {Lee},\ and\ \citenamefont {Kim}}]{Han2023}%
  \BibitemOpen
  \bibfield  {author} {\bibinfo {author} {\bibfnamefont {S.}~\bibnamefont
  {Han}}, \bibinfo {author} {\bibfnamefont {H.-W.}\ \bibnamefont {Ko}},
  \bibinfo {author} {\bibfnamefont {J.~H.}\ \bibnamefont {Oh}}, \bibinfo
  {author} {\bibfnamefont {H.-W.}\ \bibnamefont {Lee}}, \bibinfo {author}
  {\bibfnamefont {K.-J.}\ \bibnamefont {Lee}},\ and\ \bibinfo {author}
  {\bibfnamefont {K.-W.}\ \bibnamefont {Kim}},\ }\bibfield  {title} {\bibinfo
  {title} {Orbital pumping incorporating both orbital angular momentum and
  position},\ }\href {http://arxiv.org/abs/2311.00362} {\bibfield  {journal}
  {\bibinfo  {journal} {arXiv:2311.00362v1}\ } (\bibinfo {year}
  {2023})}\BibitemShut {NoStop}%
\bibitem [{\citenamefont {Pezo}\ \emph {et~al.}(2024)\citenamefont {Pezo},
  \citenamefont {Go}, \citenamefont {Mokrousov}, \citenamefont {Jaffrès},\
  and\ \citenamefont {Manchon}}]{Pezo2024}%
  \BibitemOpen
  \bibfield  {author} {\bibinfo {author} {\bibfnamefont {A.}~\bibnamefont
  {Pezo}}, \bibinfo {author} {\bibfnamefont {D.}~\bibnamefont {Go}}, \bibinfo
  {author} {\bibfnamefont {Y.}~\bibnamefont {Mokrousov}}, \bibinfo {author}
  {\bibfnamefont {H.}~\bibnamefont {Jaffrès}},\ and\ \bibinfo {author}
  {\bibfnamefont {A.}~\bibnamefont {Manchon}},\ }\bibfield  {title} {\bibinfo
  {title} {Adiabatic spin and orbital pumping in metallic heterostructures},\
  }\href {http://arxiv.org/abs/2411.13319} {\bibfield  {journal} {\bibinfo
  {journal} {arXiv.2411.13319}\ } (\bibinfo {year} {2024})}\BibitemShut
  {NoStop}%
\bibitem [{\citenamefont {Hayashi}\ \emph {et~al.}(2024)\citenamefont
  {Hayashi}, \citenamefont {Go}, \citenamefont {Haku}, \citenamefont
  {Mokrousov},\ and\ \citenamefont {Ando}}]{Hayashi2024}%
  \BibitemOpen
  \bibfield  {author} {\bibinfo {author} {\bibfnamefont {H.}~\bibnamefont
  {Hayashi}}, \bibinfo {author} {\bibfnamefont {D.}~\bibnamefont {Go}},
  \bibinfo {author} {\bibfnamefont {S.}~\bibnamefont {Haku}}, \bibinfo {author}
  {\bibfnamefont {Y.}~\bibnamefont {Mokrousov}},\ and\ \bibinfo {author}
  {\bibfnamefont {K.}~\bibnamefont {Ando}},\ }\bibfield  {title} {\bibinfo
  {title} {Observation of orbital pumping},\ }\href
  {https://doi.org/10.1038/s41928-024-01193-1} {\bibfield  {journal} {\bibinfo
  {journal} {Nature Electronics}\ }\textbf {\bibinfo {volume} {7}},\ \bibinfo
  {pages} {646} (\bibinfo {year} {2024})}\BibitemShut {NoStop}%
\bibitem [{\citenamefont {Brataas}\ \emph {et~al.}(2012)\citenamefont
  {Brataas}, \citenamefont {Tserkovnyak}, \citenamefont {Bauer},\ and\
  \citenamefont {Kelly}}]{Brataas2012b}%
  \BibitemOpen
  \bibfield  {author} {\bibinfo {author} {\bibfnamefont {A.}~\bibnamefont
  {Brataas}}, \bibinfo {author} {\bibfnamefont {Y.}~\bibnamefont
  {Tserkovnyak}}, \bibinfo {author} {\bibfnamefont {G.~E.~W.}\ \bibnamefont
  {Bauer}},\ and\ \bibinfo {author} {\bibfnamefont {P.~J.}\ \bibnamefont
  {Kelly}},\ }\bibfield  {title} {\bibinfo {title} {Spin pumping and spin
  transfer},\ }\href {https://arxiv.org/abs/1108.0385} {\bibfield  {journal}
  {\bibinfo  {journal} {arXiv preprint arXiv:1108.0385v3}\ } (\bibinfo {year}
  {2012})}\BibitemShut {NoStop}%
\end{thebibliography}%

\end{document}